\documentclass[sigconf]{acmart}

\usepackage{subcaption}
\usepackage{multirow}
\usepackage{xcolor}

\AtBeginDocument{%
  }

\setcopyright{acmlicensed}
\copyrightyear{xx}
\acmYear{xx}
\acmDOI{x.x}

\acmConference[Conf]{xx}{YY}{NY}
\acmBooktitle{ACM}
\acmISBN{xx}

\begin{document}

\title[The Mental World of LLMs in Recommendation: A Benchmark]{The Mental World of Large Language Models in Recommendation: A Benchmark on \\ Association, Personalization, and Knowledgeability}

\author{Guangneng Hu \\Xidian Univeristy}

\email{njuhgn@gmail.com}

\renewcommand{\shortauthors}{G. Hu}



\begin{abstract}
  Large language models (LLMs) have shown potential in recommendation systems (RecSys) by using them as either knowledge enhancer or zero-shot ranker. A key challenge lies in the large semantic gap between LLMs and RecSys where the former internalizes language world knowledge while the latter captures personalized world of behaviors. Unfortunately, the research community lacks a comprehensive benchmark that evaluates the LLMs over their limitations and boundaries in RecSys so that we can draw a confident conclusion. To investigate this, we propose a benchmark named LRWorld containing over 38K high-quality samples and 23M tokens carefully compiled and generated from widely used public recommendation datasets. LRWorld categorizes the mental world of LLMs in RecSys as three main scales (association, personalization, and knowledgeability) spanned by ten factors with 31 measures (tasks). Based on LRWorld, comprehensive experiments on dozens of LLMs show that they are still not well capturing the deep neural personalized embeddings but can achieve good results on shallow memory-based item-item similarity. They are also good at perceiving item entity relations, entity hierarchical taxonomies, and item-item association rules when inferring user interests. Furthermore, LLMs show a promising ability in multimodal knowledge reasoning (movie poster and product image) and robustness to noisy profiles. None of them show consistently good performance over the ten factors. Model sizes, position bias, and more are ablated.
\end{abstract}



\received{xx}
\received[revised]{xx}
\received[accepted]{xx}

\maketitle

\section{Introduction}
We humans have a mental model of the world in recommendation to make decisions and actions based on these internal and conceptual representations of the environment~\cite{ha2018world,ha2018recurrent}. For example, when someone is shopping a {\it HDMI Display Cable}, they will usually buy {\it USB 2.0 Extension Cable} as well. Another example, when someone is seeing a movie poster of {\it Kill Bill 2}, they can read the information about it knowing that the director is Quentin Tarantino, starring Uma Thurman and David Carradine, and the genres of action and adventure by using their world knowledge. They will watch this movie if these extra inferred world knowledge beyond the single poster image matches their information needs. Large language models (LLMs) recently demonstrate a great ability to internalize world knowledge and answer natural language questions in dialogue use~\cite{achiam2023gpt,touvron2023llama}, with wide applications ranging from information retrieval~\cite{qin2023large,zhu2023large} and recommendation~\cite{geng2022recommendation,zheng2023adapting,wang2024can,kang2023llms} to knowledge graph~\cite{sun2023head,sun2024large} and natural language processing~\cite{min2023recent,pan2024unifying}. Whether LLMs can be said to understand the physical and social situations that language encodes in human-like sense and without making claims on how LLMs reason internally~\cite{mitchell2023debate,shanahan2024talking,wang2024survey,huang2022language}, we investigate in a key domain which involves heavy personalized human behaviors to see if LLMs show evidence of coherent internal representations of recommendation, analogous to human mental models as shown in the above two examples.

\begin{figure*}[bt!]
  \centering
  \includegraphics[width=0.82\linewidth]{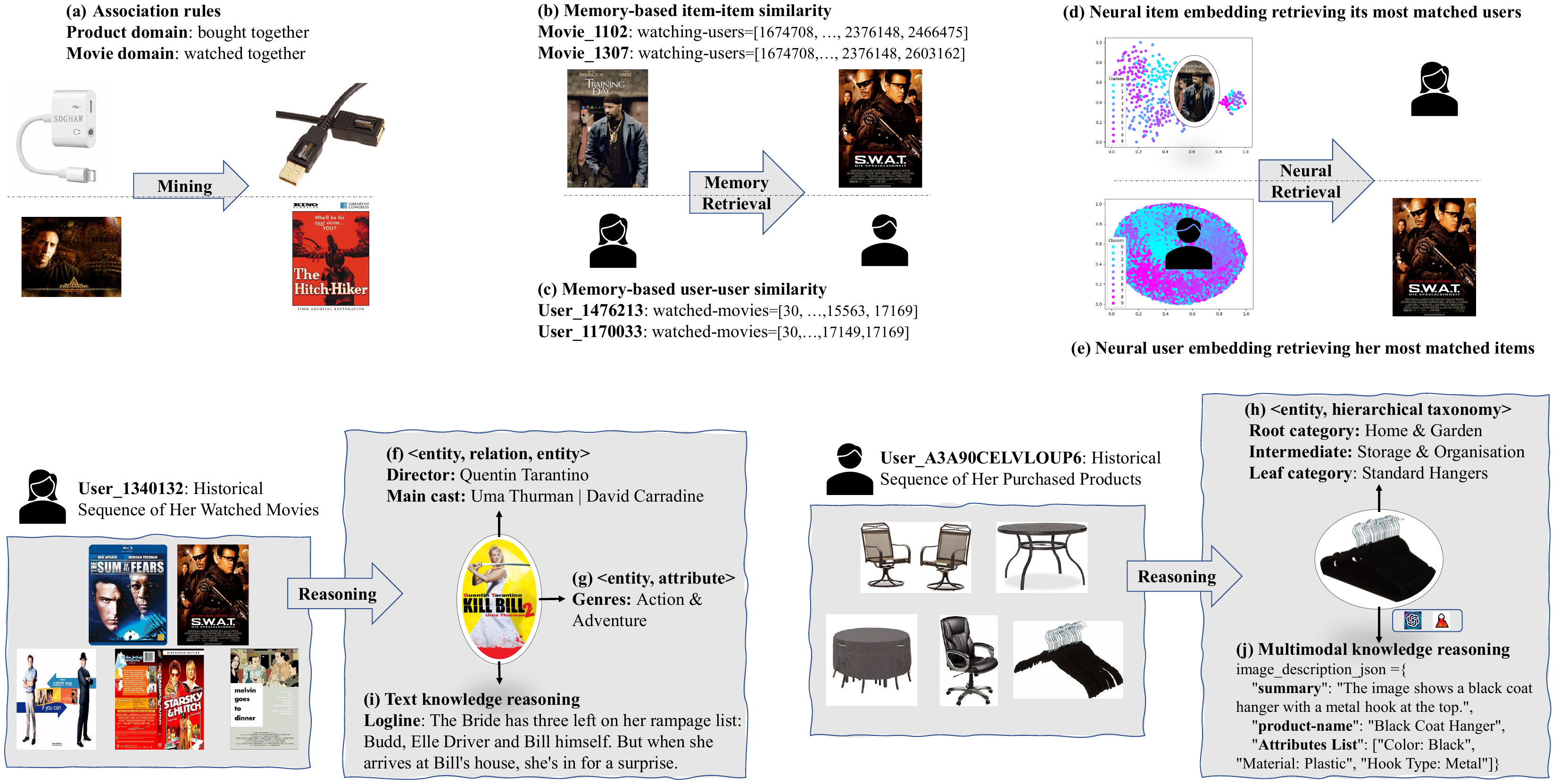}
  \caption{Measuring Mental World of LLMs in RecSys from Three Scales with 10 Factors. Part I. Association: (a) rule mining. Part II. Personalization: (b) memory-based item-item similarity, (c) memory-based user-user similarity, (d) neural item embedding retrieving its most matched user embeddings, and (e) neural user embedding retrieving her most matched item embeddings. Part III. Knowledgeability: (f) <entity, relation, entity>, (g) <entity, attribute>, (h) <entity, (hierarchical) taxonomy>, (i) text knowledge reasoning, and (j) multimodal knowledge reasoning. See Table~\ref{tab:scales-factors-description} for the detailed description of factors.}
  \label{fig:illustration-factors}
\end{figure*}

To investigate this, we delve into an unexplored area of evaluating the mental world of LLMs in RecSys by proposing a benchmark named LRWorld. It pictures the mental world of LLMs in RecSys as three main scales spanned by ten factors with 31 measures in total, as shown in Figure~\ref{fig:illustration-factors} and Table~\ref{tab:scales-factors-description}. The three key scales (association, personalization, and knowledgeability) are selected for wide coverage on what a real-world RecSys model must have the corresponding ability. The ten factors (including the two of association patterns and knowledge reasoning as shown in the above examples) mimic how humans make their decisions and actions in personalized environments using their world knowledge to satisfy their own information needs. The 31 measures (tasks) are compiled and generated from widely-used recommendation datasets to comprehensively evaluate various state-of-the-art (SOTA) LLMs families through carefully designed promptings to quantitatively picture their concrete worlds in recommendation. Starting from the LRWorld, we subsequently propose a framework for quantifying the mental worlds of LLMs in RecSys, consisting of the following three steps. i) Build the test examples to be fed into the LLMs for fully investigating their ability in various RecSys scales. ii) Design the prompt templates to instruct the LLMs for answering and probing their mental worlds in various RecSys factors. iii) Measure LLMs’ responses to capture their differences in various RecSys tasks. We make the following contributions and findings.

We make the following contributions.

\begin{itemize}
\item We are the first to establish the concept of mental world of LLMs in RecSys and conduct a pioneering evaluation on dozens of SOTA LLMs families under three key scales of association, personalization, and knowledgeability.
\item We collect a comprehensive and diverse dataset of over 38K samples and 23M tokens encompassing 10 distinct factors to fully picture the mental world of LLMs in RecSys.
\item We design and implement a testing framework over 31 measures (tasks) and 14 fields with two types of prompting templates to assess the mental world of LLMs in RecSys.
\end{itemize}

We have the following findings.
\begin{itemize}
\item We show that SOTA LLMs are good at: i) capturing association rules from users' historical sequences with HitRatio@1 of 75\%; ii) inferring entity-relation and entity-attribution of the next-item that the user will like with accuracy of 78\%, especially named entities of organizations (e.g., brand) and persons (e.g., directors, starring roles). 
\item We show that SOTA LLMs are competitive in: i) reasoning about texts (e.g., movie logline) and multimodal (e.g., product image and movie poster) knowledge with accuracy of 55\%; ii) retrieving shallow memory-based item-item similarity with HitRatio@1 of 55\%.
\item We show that SOTA LLMs are poor at: i) retrieving shallow memory-based user-user similarity with HitRatio@1 of only 29\%; ii) ranking deep embedding-based user-item matchedness with HitRatio@1 of only 13\%.
\item We further demonstrate that LLMs are generally robust to the noisy and fake user profiles over the 10 factors except for the factor of multimodal knowledge. The larger sizes of LLMs have a larger drop than their smaller counterparts, e.g., Qwen2-72B by 21\% while Qwen2-7B by 13\%. GPT-4o-mini drops by 29\%.
\item We further analyze advanced promptings. For chain-of-thought setting, it almost has no effect on the larger LLMs (Llama3-70B \& Qwen2-72B) on the factor of text knowledge reasoning. As for few-shot setting, it benefits the smaller LLMs (Llama3-8B \& Qwen2-7B) but makes their larger counterparts unstable. It almost has no effect on GPT-4o-mini.
\item We further observe that the larger LLMs do not achieve big improvement over their smaller counterparts among 8 factors. Occasionally, they degrade, e.g., Llama3.1-70B dropping by 2.7\% over Llama3.1-8B on the factor of neural embedding retrieval (LRWorld-Amazon dataset).
\end{itemize}

\begin{table*}[tb!]
\caption{Description about the Three Scales and Ten Factors of LRWorld with 31 Measures in Total. Amazon belongs to the product domain while MovieLens \& Netflix belong to movie domain. See Figure~\ref{fig:prompts-list} for the full detailed prompts for each factor and measure used to evaluate the mental world of LLMs in RecSys. See Figure~\ref{fig:benchmark-statistics} for the detailed samples statistics.}
\label{tab:scales-factors-description}
\resizebox{1.02\textwidth}{!}{
\begin{tabular}{l|l|c|l}
\toprule
Scale           & \multicolumn{1}{c|}{Factor}                                                                          & Number & \multicolumn{1}{c}{Description (each factor corresponding to several numbers of measures)}                                                                                                                                                            \\ \midrule
Association      & \multicolumn{1}{l|}{association rules mining}                                                                            & 3       & \begin{tabular}[c]{@{}l@{}}Amazon: mine the products which are frequently purchased together in users' transactions. \\ MovieLens \& Netflix: mine the movies which are frequently watched in users' history.  \end{tabular}                                                                                                                    \\ \hline
\multicolumn{1}{l|}{\multirow{7}{*}{Personalization}}  & \begin{tabular}[c]{@{}l@{}}memory-based\\ item-item similarity\end{tabular}                          & 3       & \begin{tabular}[c]{@{}l@{}}Amazon: find the simiar products by their shared co-purchasing users. \\ MovieLens \& Netflix: find the simiar movies by their shared co-watching users. \end{tabular}                                                                                                                    \\
\cline{2-4}
                 & \begin{tabular}[c]{@{}l@{}}memory-based\\ user-user similarity\end{tabular}                          & 3       & \begin{tabular}[c]{@{}l@{}}Amazon: find the simiar users by their shared co-purchased products. \\ MovieLens \& Netflix: find the simiar users by their shared co-watched movies. \end{tabular}                                                                                                                    \\
\cline{2-4}
                 & \begin{tabular}[c]{@{}l@{}}neural item embedding\\ retrieving its matched users\end{tabular}         & 3       & \begin{tabular}[c]{@{}l@{}}Amazon: retrieve matched users for the targeted product by their cosine similarity of neural embeddings. \\ MovieLens \& Netflix: retrieve matched users for the targeted movie by their cosine similarity of neural embeddings. \end{tabular}                                                                                                                    \\
\cline{2-4}
                 & \begin{tabular}[c]{@{}l@{}}neural user embedding\\ retrieving her matched items\end{tabular}         & 3       & \begin{tabular}[c]{@{}l@{}}Amazon: retrieve matched products for the targeted user by their cosine similarity of neural embeddings. \\ MovieLens \& Netflix: retrieve matched movies for the targeted user by their cosine similarity of neural embeddings. \end{tabular}                                                                                                                    \\ \hline
\multicolumn{1}{l|}{\multirow{7}{*}{Knowledgeability}}   & \textless{}entity, relation, entity\textgreater{}                                                    & 5       & \begin{tabular}[c]{@{}l@{}} Amazon: (\textless{}product,brand\textgreater{}) infer the brand of the product that the user will buy next, instead of the product itself.\\  MovieLens \& Netflix: (\textless{}movie,cast\textgreater{}, \textless{}movie,director\textgreater{}) infer the starring roles and directors of the movie \\ $\quad$ that the user will watch next, instead of the movie itself.\end{tabular}    \\
\cline{2-4}
                 & \textless{}entity, attribute\textgreater{}                                                           & 3       & \begin{tabular}[c]{@{}l@{}}Amazon: (\textless{}product,color\textgreater) infer the color of the product that the user will buy next, instead of the product itself.\\ MovieLens \& Netflix: (\textless{}movie,genres\textgreater{}) infer the genres of the movie that the user will watch next, instead of the movie itself. \end{tabular} \\
\cline{2-4}
                 & \begin{tabular}[c]{@{}l@{}}\textless{}entity, \\ (hierarchical) taxonomy\textgreater{}\end{tabular} & 3       & \begin{tabular}[c]{@{}l@{}} Amazon: (\textless{}product,leaf\textgreater{}, \textless{}product,intermediate\textgreater{}, \textless{}product, root\textgreater{})  infer the hierarchical taxonomies \\ $\quad$ from leaf to intermediate to root of the product that the user will buy next, instead of the product itself. \end{tabular}                           \\
\cline{2-4}
                 & text knowledge reasoning   & 2    & \begin{tabular}[c]{@{}l@{}} MovieLens \& Netflix: (\textless{}movie,logline\textgreater{}) infer the logline of the movie that the user will watch next, instead of the movie itself. \end{tabular}  \\
\cline{2-4}
                 & multimodal knowledge                                                                                 & 3       & \begin{tabular}[c]{@{}l@{}}Amazon: product images and descriptions generated by large multimodal models (LMMs). \\ MovieLens \& Netflix: movie posters and descriptions generated by LMMs. \end{tabular}   \\  
\bottomrule   
\end{tabular}
}
\end{table*}

\section{Measuring Mental World of LLMs in RecSys}

We define the three scales and ten factors with 31 measures to picture the mental world of LLMs in RecSys in this section. The illustrations are shown in Figure~\ref{fig:illustration-factors} and the descriptions are shown in Table~\ref{tab:scales-factors-description} respectively. We will describe how we build the benchmark dataset for instantiating them in the next section. 

Here, besides existing works mainly focusing on evaluating the rating prediction or binary preference and aspects of items~\cite{kang2023llms,sanner2023large,yoon2024evaluating}, we go further to categorize the mental world of LLMs in RecSys from three key scales---association, personalization, and knowledgeability---and ten factors. It widely covers rules and patterns mining, traditional memory-based item-item (user-user) similarity and deep neural embeddings retrieval, entity-relation and entity-attribute recognition, hierarchical taxonomy capturing, and texts and multimodal knowledge reasoning.

\subsection{Association: Rules Mining}

Association rules mining is to discover and identify the rules that determine how and why certain items are connected~\cite{agrawal1993mining}. For example, when someone is shopping a HDMI Display Cable, they will usually buy USB 2.0 Extension Cable as well, in real-word transactions of Amazon.com e-commerce platforms. Such rules can be used as the basis for making decisions about products' promotion, placements, and recommendations~\cite{lin2002efficient}. As a result, we consider the {\it Association} scale as a key ability that LLMs must have in the mental world of RecSys, as shown in Figure~\ref{fig:illustration-factors}a. 

\subsection{Personalization: Memory-based and Neural Embeddings Similarity Retrieving}

Personalization requires recommender systems to provide personalized items for individual users to satisfy their own information needs according to their different profiles, context, and behaviors~\cite{ricci2021recommender}. The traditional memory-based collaborative filtering and the deep learning-based neural embeddings are two main ways of RecSys to achieve the personalization services~\cite{linden2003amazon,he2017neural}. As a result, we consider the {\it Personalization} scale as a key ability that LLMs must have in the mental world of RecSys, as shown in Figure~\ref{fig:illustration-factors}(b-e). 

\noindent
{\bf Memory-based similarity retrieving} Memory-based collaborative filtering uses the stored user-item ratings to directly predict preferences for new items~\cite{linden2003amazon}. It can be done in two ways known as item-based and user-based recommendation. For the {\bf item-based} recommendation, the core is to find the neighbors which have similar characteristics with the target item by computing their co-rating users. As shown in Figure~\ref{fig:illustration-factors}b, the two movies ({\it Training Day (2001)} and {\it S.W.A.T. (2003)} in the Netflix dataset) are similar because they have large overlapped set of co-rating users. Likewise, for the {\bf user-based} recommendation, the core is to find the neighbors who have similar interests with the target user by computing their co-rated items. As shown in Figure~\ref{fig:illustration-factors}c, the two users are similar because they have large overlapped set of co-rated movies. Given the target item (and user), we require the LLMs to correctly retrieve the similar items (and users) from a candidate pool. 

\noindent
{\bf Neural embeddings similarity retrieving} In contrast to the shallow memory-based similarity, the neural collaborative filtering learns the deep embeddings for users and items by neural networks and then computes their matching degree based on these learned embedding representation. As shown in Figure~\ref{fig:illustration-factors}d, the target movie ({\it Training Day (2001)} in the Netflix dataset) is represented as a point in the embedding space, and we require the LLMs to correctly {\bf retrieve the matched users} (also represented as data points in the same embedding space) from a candidate pool by their distances. Likewise, Figure~\ref{fig:illustration-factors}e shows LLMs to correctly {\bf retrieve the matched items} for the target item in the embedding space. Note, the embedding space is firstly learned by an off-the-shell neural matrix factorization method~\cite{he2017neural} so that we know the ground truth about which items (users) are matched to the target user (item). As we will see in the later experiments, this is the most challenging task for LLMs to be successful in replacing the recommendation models. This is also the big semantic gap between LLMs and RecSys since the former internalizes language world knowledge while the latter captures collaborative personalized behaviors. 

\subsection{Knowledgeability: Knowledge Graph, Taxonomy, and Multimodal Reasoning}\label{paper:knowledge}

RecSys suffers from sparsity and long-tail issues~\cite{zhang2021model,wu2024coral}. Knowledge graph-enhanced RecSys is an effective approach for accurate and explainable recommendation~\cite{guo2020survey,wang2018ripplenet,zhao2017meta,hu2018leveraging}. Large-scale knowledge graphs deployed in the e-commerce platforms like Amazon.com can improve customers' shopping experience and engagement and achieve maximum revenues in return~\cite{yu2024cosmo}. As a result, we consider the {\it Knowledgeability} scale as a key ability that LLMs must have in the mental world of RecSys, as shown in Fig~\ref{fig:illustration-factors}(f-g). Actually, LLMs are very good at commonsense knowledge which can be used for capturing semantic relationships among items, deriving user intentions, and providing explainable recommendations.

\begin{figure}[bt!]
  \centering
  \includegraphics[width=1.05\linewidth]{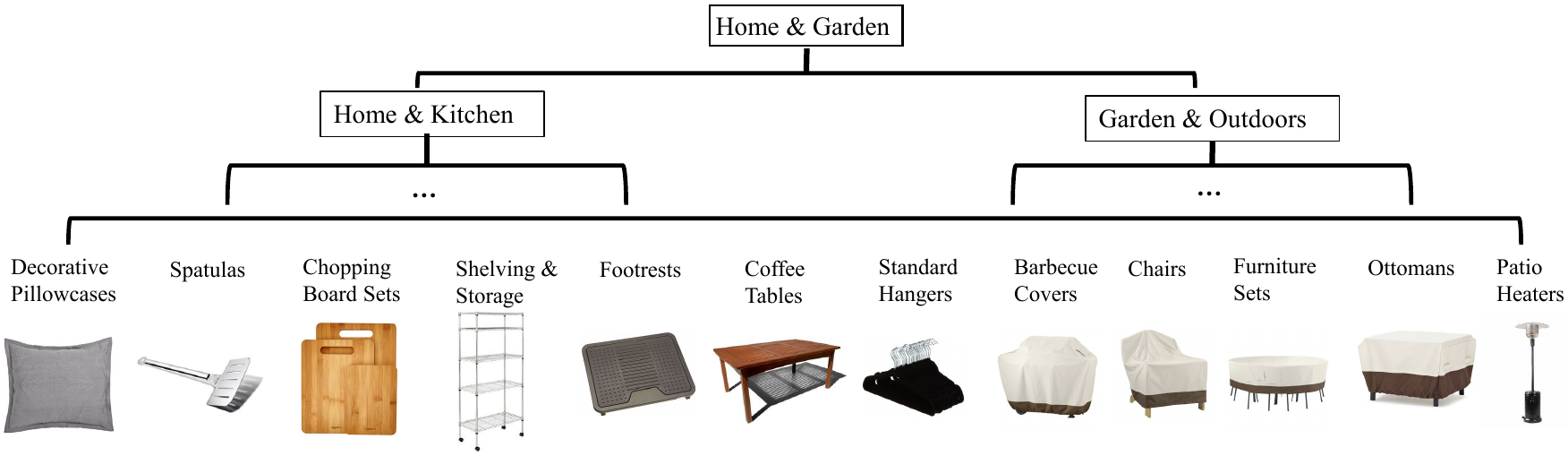}
  \caption{Illustrations on Taxonomy with Root Category ``Home \& Garden'' from Our LRWorld (Amazon) Benchmark. The last row is products corresponding to each of the 12 leaf categories. There is 7-layer depth of the hierarchical taxonomy for this root and only one intermediate layer is shown which are {\it Home \& Kitchen} and {\it Garden \& Outdoors}.}
  \label{fig:taxo}
\end{figure}

\begin{figure}[bt!]
  \centering
  \includegraphics[width=1.05\linewidth]{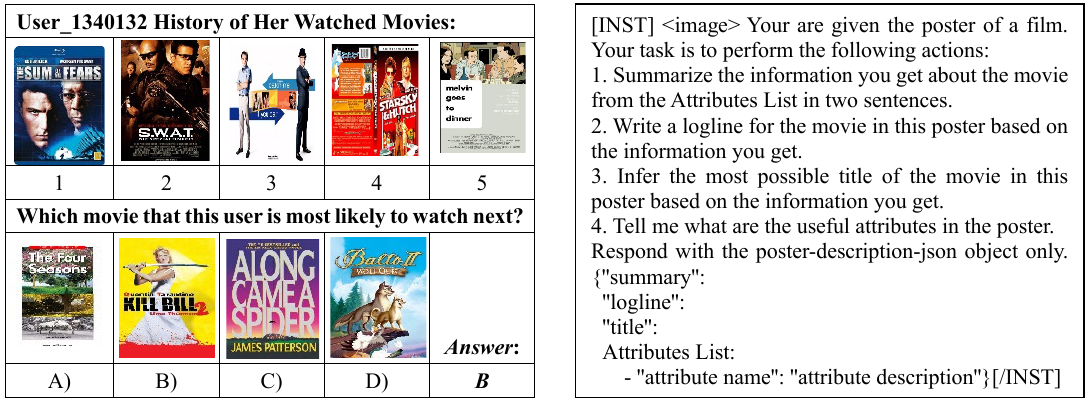}
  \caption{Left: Illustrations on Multimodal Knowledge Reasoning about Posters from Our LRWorld (Netflix) Benchmark. Right: Prompts of Instructing Large Multimodal Models to Generate Detailed Text Description for Poster Images.}
  \label{fig:poster2desc}
\end{figure}

\begin{figure}[bt!]
  \centering
  \includegraphics[width=1.05\linewidth]{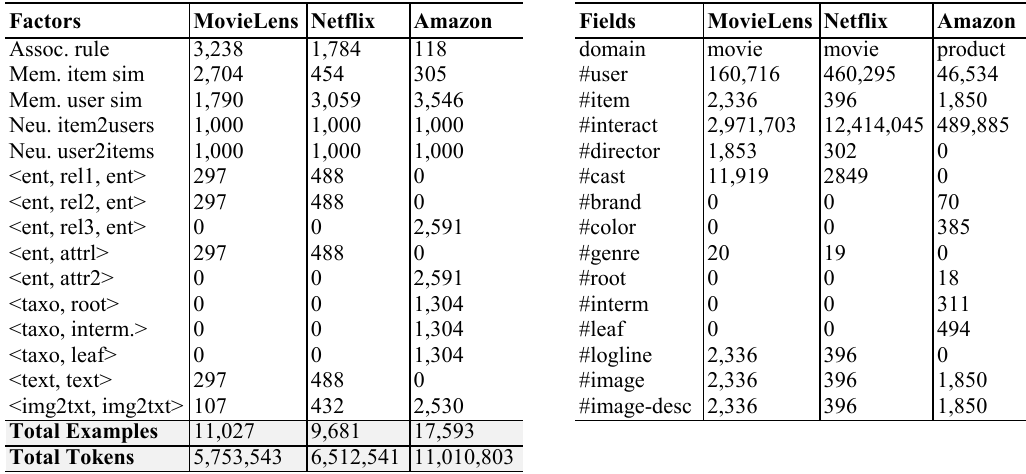}
  \caption{Statistics of Our LRWorld Benchmark. Left: Factors. Right: Fields. See Table~\ref{tab:scales-factors-description} for the detailed description of the factors and fields. The total tokens are fed into the LLMs.}
  \label{fig:benchmark-statistics}
\end{figure}

Besides existing works mainly focusing on the two typical factors of entity-relation-entity and entity-attribute ~\cite{jiang2024diffkg,yu2024cosmo,sun2024head}, we go further to picture the mental world of LLMs in RecSys on Knowledgeability scale with three more factors, including hierarchical taxonomies over products, unstructured texts (movie logline), and multimodal knowledge reasoning (movie poster/ product image).  

\noindent
{\bf <entity,relation,entity>} These tuples are the basic elements of a knowledge graph (KG) where nodes are entities and edges represent their relations. Items and users and their attributes and profiles can be mapped into the KG to waver an interconnected graph representing and propagating their mutual relations and messages. As shown in Figure~\ref{fig:illustration-factors}f, the entities are the movie ({\it Kill Bill 2}), the director (Quentin Tarantino), and the main cast (Uma Thurma and David Carradine) while the relations are {\it directed-by} and {\it acted-by} among these entities. Given the watching history of the target user, we require LLMs to predict the director and main cast of the movie that the user is likely to watch next. This factor measures the reasoning ability of LLMs to find correlations not only between historical movies and the next-movie but also between the movie-director and movie-cast relations among these entities.

\noindent
{\bf <entity,attribute>} As shown in Figure~\ref{fig:illustration-factors}g, this factor is similar to the above  <entity,relation,entity> one but focusing on the item's attributes instead of item relations, e.g., the genre of the movie, the color of the product, and the brand of the product et al.

\noindent
{\bf Hierarchical taxonomy} Different from the knowledge graph, the taxonomy is a meta knowledge about the knowledge graph. As shown in Figure~\ref{fig:illustration-factors}h, the product {\it Black Hangers} has a hierarchical taxonomy of it belonging to the leaf category {\it Standard Hangers}, intermediate category {\it Storage \& Organisation}, and the root category {\it Home \& Garden} under the Amazon.com taxonomies. See Figure~\ref{fig:taxo} for a more complete hierarchical taxonomy. Given the purchasing history of the target user, we require LLMs to predict the hierarchical categories of the product that the user is likely to buy next. This factor measures the reasoning ability of LLMs to find correlations not only between the historical product and the to-buy product but also between the hierarchical taxonomy among these categories from root to intermediate to leaf nodes. In other words, LLMs need to know that {\it Standard Hangers} products are a type of the {\it Storage \& Organisation}, and they are also a type of the {\it Home \& Garden}.

\noindent
{\bf Text knowledge reasoning} LLMs are good at understanding and generating unstructured texts. As shown in Figure~\ref{fig:illustration-factors}i and Table~\ref{tab:illustration-logline}, given a target user's history of her latest watched movies, we require the LLMs to predict the logline of the movie that this user will watch next. Note, the logline is provided only without movie title. The LLMs need to implicitly build the user profile by inferring the movie's title, director, starring roles, genres, and intentions from these loglines, so as to match the candidate movie correctly. It is not easy for humans to comprehend these loglines for achieving the right answer. Have a try the example of Table~\ref{tab:illustration-logline}.

\noindent
{\bf Multimodal knowledge reasoning} Visual features are a key to understand movies and hence to better capture users' preferences~\cite{zhao2016matrix}.  As shown in Figure~\ref{fig:poster2desc}, given a target user's history of her latest watched movies, we require LLMs to predict the poster of the movie that this user will watch next. Note, the poster image is provided only without any other information (like movie title, genre, director, and starring roles). Considering the existing large multimodal models (LMMs) mainly support single image input, we represent the user's sequence of movie posters as their detailed text descriptions extracted by LMMs (see Figure~\ref{fig:poster2desc} for prompts). The details will be described in the next section.


\section{Benchmark Construction}\label{paper:benchmark}

We build LRWorld from five public sources covering different domains and diverse characteristics. Statistics are shown in Fig~\ref{fig:benchmark-statistics}.

\noindent
{\bf Original sources and data alignment} (i) {\bf LRWorld (Amazon)} We start from the {\bf Amazon Review} (2018) dataset~\cite{dataset-amazon-ucsd}, specifically the 5-core subsets~\cite{ni2019justifying}. To get the product images, we align it with {\bf Amazon Berkeley Objects} dataset~\cite{dataset-amazon-berkeley} by ASINs (Amazon Standard Identification Number) which has 398,212 unique catalog images~\cite{collins2022abo}. (ii) {\bf LRWorld (Netflix)} We start from the {\bf Netflix KDD-Cup} 2007 dataset~\cite{dataset-Netflix}. It has 100 million ratings with 17 thousand movies by 480 thousand users. To get the meta information~\cite{dataset-kaggle} and the poster images of the movies, we align it with the recent work {\bf LLMRec}~\cite{dataset-hkuds} by the movie name and release year~\cite{wei2024llmrec}. (ii) {\bf LRWorld (MovieLens)} We start from the {\bf MovieLens 25M} dataset~\cite{dataset-movielens}. It has 25 million ratings with 62,000 movies by 162,000 users. Similar to the above LRWorld (Netflix), to get the meta information and the poster images of the movies, we align it by the movie name and release year. Note, the user set and rating behaviors are very different among Netflix and MovieLens though they are both belonging to the movie domain.

\noindent
{\bf Constructing ground truth and positives} For the ten factors (see Table~\ref{tab:scales-factors-description}), the positives are constructed in the following ways. (a) {\bf Association rules}: The users' historical sequences are considered as the transactions in market analysis. We use the Apriori algorithm~\cite{agrawal1994fast} implemented in the Python Mlxtend package to get the ground truth association rules. (b) {\bf Memory-based item-item similarity}: The positives are the <item,item> pairs computed by high cosine similarity of their one-hot vectors where the element is one if someone watched/purchased it. (c) {\bf Memory-based user-user similarity}: It is similar to (b). (d) {\bf Neural item embedding retrieving its matched users}: The positives are the <target item, matched users> pairs computed by high cosine similarity of their embedding vectors learned by neural matrix factorization~\cite{he2017neural}. To be diverse and representative, the pool of target items are the centroid prototypes of k-means clustering over the embeddings. This is challenging for LLMs since these embeddings encode high-order collaborative behaviors learned by traditional RecSys from rating matrix. (e) {\bf Neural user embedding retrieving her matched items}: It is similar to (d).  (f) {\bf <entity,relation,entity>}: Given a user's sequence over items, say, $(i_1,i_2,i_3,i_4,i_5,i_6)$, the latest item $i_6$ is called positive item of its previous history $(i_1,i_2,i_3,i_4,i_5)$ in a sliding window way. This is the widely-used leave-one-out evaluation strategy. After we determine the positive item, we treat it as the head-entity. Then we ask LLMs to predict the tail-entity for each different relation. For example, we ask LLMs to predict the tail-entity=director of the head-entity=movie for the relation={\it directed-by}. (g) {\bf <entity,attribute>}: Similar to the (f), we instead we ask LLMs to predict the attribute of the positive item. For example, we ask LLMs to predict the attribute=color of the entity=product.  (h) {\bf <entity,hierarchical taxonomy>}: Similar to the (g), we instead we ask LLMs to predict the hierarchical taxonomy of the positive item from root to intermediate to leaf categories. For example, we ask LLMs to predict the taxonomy=\{root,intermediate,leaf\}-categories of the entity=product (see Figure~\ref{fig:taxo}). (i) {\bf Text knowledge reasoning}: Similar to the (f), the user's sequence over item indices $(i_1,i_2,i_3,i_4,i_5,i_6)$ is now replaced by the item texts $(t_1,t_2,t_3,t_4,t_5,t_6)$ where $t$ is the item text for the corresponding item index $i$. The item text is the movie logline (one-sentence summary) (see Table~\ref{tab:illustration-logline} for an example). (j) {\bf Multimodal knowledge reasoning}: Similar to the (f), the user's sequence over item indices $(i_1,i_2,i_3,i_4,i_5,i_6)$ is now replaced by the item's image $(v_1,v_2,v_3,v_4,v_5,v_6)$ where $v$ is the item image for the corresponding item index $i$. The item image is the movie poster or product image. Considering the existing large multimodal models (LMMs) mainly support single image input, we instead represent the user's sequence of item images as their detailed text descriptions which are firstly extracted by LMMs. Specifically, we instruct the LLaVA-1.6 model to extract text description from image~\cite{liu2024visual,liu2024llavanext}. See Figure~\ref{fig:poster2desc} for the fine-grained extraction prompt and an example.

\section{Prompting Design}\label{paper:prompt-design}

The full evaluation prompting templates are listed in Figure~\ref{fig:prompts-list}. The few-shot evaluation prompt is shown in Figure~\ref{fig:prompts-fewshot-logline}. 

\noindent
{\bf Two types of templates} In order to fully picture LLMs' mental world in RecSys and take into account the characteristics of different measuring factors, we designed two types of templates covering both simple-formed Multiple-Choice Question (MCQ) over 4 candidates and difficult HitRate@1 over 10 candidates. Specifically, the templates are the type of HitRate@1 for the five factors of {\it Association} and {\it Personalization}, while the templates are the type of MCQ for the five factors of {\it Knowledgeability}. See Table~\ref{tab:illustration-logline} and Table~\ref{tab:example-user2matchedItems} as an example for the type of MCQ and HitRate@1 respectively.  

\noindent
{\bf Question generation} To probe the mental world of LLMs in RecSys, we need to feed test examples into them. Besides the positives described in Section~\ref{paper:benchmark}, we need to insert negatives into the prompt templates so as to get complete test examples. For the type of MCQ template, we randomly sample three negatives per one positive. See Figure~\ref{fig:poster2desc} as an example, the choice B is the positive while all other choices are the three negatives. For the type of HitRate@1 template, we randomly sample nine negatives per one positive. See Table~\ref{tab:example-user2matchedItems} as an example, the index 4 is the positive while all other choices are the nine negatives. We do not purposely generate {\it hard} negatives since we have found that these random negatives are already difficult for LLMs. The prediction errors and popularity bias are ablated later. In current study, we evaluate LLMs to pick one choice from a pool of controlled candidates rather than to generate recommendations directly~\cite{li2024large}, which is left for future work. Distributions of correct answers are shown in Table~\ref{tab:distribution-Amazon}-\ref{tab:distribution-Netflix}.

\begin{table*}[tb!]
\caption{Results on LRWorld (Amazon). RANDOM is to randomly pick one choice from the candidates. See Section~\ref{exp:llm-versions} for the specific version of the LLMs. See Table~\ref{tab:scales-factors-description} for the detailed description of each scale and factor. The higher the better.} 
\label{tab:Amazon}
\resizebox{.82\textwidth}{!}{
\begin{tabular}{l|c|c|c|c|c|c|c|c|c|c|c}
\toprule
\multicolumn{1}{l|}{Scales} & \multicolumn{1}{c|}{Association} & \multicolumn{4}{c|}{Personalization}                                     & \multicolumn{6}{c}{Knowledgeability}                                                                                    \\ \midrule
Factors & \multicolumn{1}{c|}{\begin{tabular}[c]{@{}c@{}}Assoc \\ rule\end{tabular}} & \multicolumn{1}{c|}{\begin{tabular}[c]{@{}c@{}}Mem \\ item-item\end{tabular}} & \multicolumn{1}{c|}{\begin{tabular}[c]{@{}c@{}}Mem \\ user-user\end{tabular}} & \multicolumn{1}{c|}{\begin{tabular}[c]{@{}c@{}}Neu emb\\ item2users\end{tabular}} & \multicolumn{1}{c|}{\begin{tabular}[c]{@{}c@{}}Neu emb\\ user2items\end{tabular}} & \multicolumn{1}{c|}{\begin{tabular}[c]{@{}c@{}}KG \\ (prod,brand)\end{tabular}} & \multicolumn{1}{c|}{\begin{tabular}[c]{@{}c@{}}KG \\ (prod,color)\end{tabular}} & \multicolumn{1}{c|}{\begin{tabular}[c]{@{}c@{}}Taxo \\ (prod, leaf)\end{tabular}} & \multicolumn{1}{c|}{\begin{tabular}[c]{@{}c@{}}Taxo \\ (prod, interm)\end{tabular}} & \multicolumn{1}{c|}{\begin{tabular}[c]{@{}c@{}}Taxo \\ (prod, root)\end{tabular}} & \multicolumn{1}{c}{\begin{tabular}[c]{@{}c@{}}Multimodal\\ knowledge\end{tabular}} \\
\midrule 
RANDOM & 0.1000 & 0.1000 & 0.1000 & 0.1000 & 0.1000 & 0.2500 & 0.2500 & 0.2500 & 0.2500 & 0.2500 & 0.2500\\ \hline
Falcon & 0.1016 & 0.0983 & 0.1060 & 0.1030 & 0.0970 & 0.2470 & 0.2612 & 0.2331 & 0.2262 & 0.2722 & 0.2379\\ \hline
Vicuna-1.5 & 0.1694 & 0.2950 & 0.1037 & 0.1030 & 0.0990 & 0.2732 & 0.2454 & 0.2315 & 0.2124 & 0.2415 & 0.2399\\ \hline
FLAN-T5 & 0.3644 & 0.4163 & 0.1204 & {\bf 0.1180} & 0.1080 & 0.5993 & 0.3087 & 0.3788 & 0.4762 & 0.1832 & 0.3007\\ \hline
Llama-3 & 0.2372 & {\bf 0.5540} & 0.1618 & 0.0950 & 0.1060 & 0.7348 & 0.3894 & 0.5122 & 0.5866 & 0.3266 & 0.2861\\ \hline
Llama-3.1 & 0.5254	&0.5442&	0.1508&	0.0950	&0.1090	&0.7506&	0.3948&	0.5099	&0.5966&	0.3865	&0.2897 \\ \hline
Phi-3 & 0.5169 & 0.5016 & 0.1505 & 0.0900 & 0.0950 & 0.7271 & 0.4920 & 0.4953 & 0.5828 & 0.3228 & {\bf 0.5312}\\ \hline
Mistral & 0.5593 & 0.5180 & 0.1438 & 0.0920 & {\bf 0.1140} & 0.7479 & 0.5179 & 0.4762 & 0.5851 & 0.5352 & 0.2897\\ \hline
Qwen-2 & 0.4067 & 0.5114 & 0.1404 & 0.1050 & 0.0940 & 0.7383 & 0.4654 & 0.4877 & 0.5605 & 0.3082 & 0.4750\\ \hline
LlaVA-1.6 & 0.1779 & 0.3901 & 0.1446 & 0.0980 & 0.0930 & 0.7263 & 0.3720 & 0.4961 & 0.5720 & {\bf 0.5567} & 0.3442\\ \hline
GPT-3.5-Turbo & 0.4491 & 0.4622 & {\bf 0.1914} & 0.0970 & 0.0980 & 0.7190 & 0.4145 & 0.4754 & 0.5820 & 0.3489 & 0.2158\\ \hline
GPT-4o-mini & {\bf 0.7542} & 0.5508 & 0.1720 & 0.1130 & 0.1080 & {\bf 0.7803} & {\bf 0.5569} & {\bf 0.5682} & {\bf 0.6196} & 0.4432 & 0.5031\\ 
\bottomrule                                                                            
\end{tabular}
}
\end{table*}

\begin{table*}[tb!]
\caption{Results on LRWorld (MovieLens). See Table~\ref{tab:scales-factors-description} for the detailed description of each scale and factor. The higher the better.}
\label{tab:Movielens}
\resizebox{.77\textwidth}{!}{
\begin{tabular}{l|c|c|c|c|c|c|c|c|c|c}
\toprule
\multicolumn{1}{l|}{Scales} & \multicolumn{1}{c|}{Association} & \multicolumn{4}{c|}{Personalization}                                     & \multicolumn{5}{c}{Knowledgeability}                                                                                    \\ \midrule 
Factors & \multicolumn{1}{c|}{\begin{tabular}[c]{@{}c@{}}Assoc \\ rule\end{tabular}} & \multicolumn{1}{c|}{\begin{tabular}[c]{@{}c@{}}Mem \\ item-item\end{tabular}} & \multicolumn{1}{c|}{\begin{tabular}[c]{@{}c@{}}Mem \\ user-user\end{tabular}} & \multicolumn{1}{c|}{\begin{tabular}[c]{@{}c@{}}Neu emb\\ item2users\end{tabular}} & \multicolumn{1}{c|}{\begin{tabular}[c]{@{}c@{}}Neu emb\\ user2items\end{tabular}} & \multicolumn{1}{c|}{\begin{tabular}[c]{@{}c@{}}KG \\ (movie,cast)\end{tabular}} & \multicolumn{1}{c|}{\begin{tabular}[c]{@{}c@{}}KG \\ (movie,director)\end{tabular}} & \multicolumn{1}{c|}{\begin{tabular}[c]{@{}c@{}}KG \\ (movie, genre)\end{tabular}} & \multicolumn{1}{c|}{\begin{tabular}[c]{@{}c@{}}Text \\ knowledge \end{tabular}} & \multicolumn{1}{c}{\begin{tabular}[c]{@{}c@{}}Multimodal\\ knowledge\end{tabular}} \\ 
\midrule 
RANDOM & 0.1000 & 0.1000 & 0.1000 & 0.1000 & 0.1000 & 0.2500 & 0.2500 & 0.2500 & 0.2500 & 0.2500\\ \hline
Falcon & 0.1040 & 0.1061 & 0.1016 & 0.0940 & 0.0840 & 0.2861 & 0.3030 & 0.2929 & 0.2929 & 0.2523\\ \hline
Vicuna-1.5 & 0.2764 & 0.1719 & 0.1363 & 0.0940 & 0.0840 & 0.2390 & 0.2222 & 0.2457 & 0.2895 & 0.2336\\ \hline
FLAN-T5 & 0.1717 & {\bf 0.3653} & 0.1916 & 0.1070 & 0.0960 & 0.3771 & 0.5892 & 0.4579 & 0.2828 & 0.3084\\ \hline
Llama-3 & 0.3977 & 0.2851 & 0.2223 & 0.0970 & 0.1140 & 0.5084 & 0.5353 & 0.4175 & 0.3367 & 0.2242\\ \hline
Llama-3.1 &0.6602&	0.2585&	0.2424&	0.0950 &	0.1100	&0.5185	&0.4747&	0.4478&	0.3063	&0.1869 \\ \hline
Phi-3 & 0.4771 & 0.2363 & 0.2000 & 0.0890 & 0.1260 & 0.4040 & 0.6094 & 0.4040 & 0.3501 & 0.3177\\ \hline
Mistral & 0.5135 & 0.2943 & 0.2346 & 0.0920 & 0.1040 & 0.5117 & 0.6868 & 0.4377 & 0.2996 & 0.3084\\ \hline
Qwen-2 & 0.6547 & 0.2892 & 0.2379 & 0.1030 & 0.1000 & 0.5319 & 0.6060 & 0.4814 & 0.4276 & 0.2056\\ \hline
LlaVA-1.6 & 0.5463 & 0.2647 & 0.2553 & {\bf 0.1080} & 0.1010 & 0.4781 & {\bf 0.7104} & 0.3905 & 0.4579 & {\bf 0.3364}\\ \hline
GPT-3.5-Turbo & {\bf 0.6726} &  0.3472 & 0.2357 & 0.0890 & {\bf 0.1300} & 0.4747 & 0.5622 & 0.4309 & 0.2962 & {\bf 0.3364}\\ \hline
GPT-4o-mini & 0.5975 & 0.2614 & {\bf 0.2966} & 0.0940 & 0.1100 & {\bf 0.6565} & 0.6632 & {\bf 0.4949} & {\bf 0.5117} & 0.3177\\ 
\bottomrule                                                                            
\end{tabular}
}
\end{table*}

\section{Experiments}


\subsection{Experimental Setup}\label{exp:llm-versions}

\noindent
{\bf Large language models} We evaluate on dozens of state-of-the-art LLMs. See Appendix~\ref{appendix:llm-versions} and Table~\ref{tab:model-version} for details.

\noindent
{\bf Metrics} For the type of Multiple-Choice Question (MCQ) templates and questions, we compute the accuracy metric~\cite{hendrycks2020measuring,huang2024c} which measures the number of questions that LLMs pick the correct answers over the total number of all this kind of questions. For the type of HitRate@1 templates and questions, we compute the top-1 hit ratio metric~\cite{zheng2023adapting} which measures the share of hits for which the correct answers rank in the first. Missed rate statistics are in Appx~\ref{appendix:miss-rate}.

\noindent
{\bf Implementation details} All LLMs are evaluated with the same test questions of LRWorld for all of the ten factors. OpenAI GPTs are evaluated via API requests, while others are evaluated through PyTorch Transformers with Hugging Face or the released code by their authors. To obtain deterministic and reproducible responses from LLMs, temperature set to 0, except for 0.01 of OpenAI's GPTs. For association rule, the min support and confidence set to 0.01. For memory similarity, threshold to 0.01. For neural embedding, clusters are 10. To alleviate the position bias of LLMs, we randomly shuffle the order of the correct answer in the candidates~\cite{wang2023large,wettig2024qurating}.

\subsection{Mental World of LLMs in RecSys: The Good}
Results are shown in Table~\ref{tab:Amazon}, Table~\ref{tab:Movielens}, and Table~\ref{tab:Netflix}.

\noindent
{\bf LLMs can evoke a good sense of the {\it Association} and {\it Knowledgeability} scales.} In the product domain, the best LLM achieves a HitRatio@1 score of 0.7542 on the Association scale, while it is 0.6726 in the movie domain. In terms of the Knowledgeability scale, the best LLM achieves an accuracy of 0.7803 in the product domain, while it is 0.7104 in the movie domain. 

\noindent
{\bf LLMs are good at capturing the named entity of the {\it organization} and {\it person} factors.} In the product domain, the best LLM achieves an accuracy of 0.7803 on the <product,brand> factor. In the movie domain, the best LLM achieves an accuracy of 0.7104 on the <movie,director> factor, and 0.6565 on <movie,cast>. 

\noindent
{\bf LLMs show equally well ability in both {\it text} knowledge and {\it multimodal} knowledge reasoning.} In the product domain, the best LLM achieves an accuracy of 0.5312 on <product,image2text> multimodal knowledge. In the movie domain, the best LLM achieves an accuracy of 0.5117 on the <movie,logline> text knowledge. Recall that, the multimodal knowledge is firstly extracted using a large multimodal model as shown in Figure~\ref{fig:poster2desc}. 

\subsection{Mental World of LLMs in RecSys: The Bad}
\noindent
{\bf LLMs are struggling in picturing the world of {\it Personalization} scale.} In general, we observe an overall decreasing trend in HitRatio@1 for LLMs from the memory-based to neural embedding factors on both the product and movie domains. Recall that, memory-based similarity is shallow user-user (item-item) relations computed by their one-hot vectors, while neural embedding similarity is nonlinear high-order relations learned by the neural matrix factorization~\cite{he2017neural}. In specific, the best LLM achieves a HitRatio@1 score of 0.5540 on the memory-based factors in the product domain, while it is 0.1300 on the neural embedding factors in the movie domain. This might be due to that LLMs need to be better aligned with user preferences and relevant context~\cite{lyu2024llm}.

\noindent
{\bf LLMs are more struggling in capturing the {\it user} factors than {\it item} factors.} On all the three LRWorld-Amazon/MovieLens/Netflix datasets, LLMs are more struggling in user-user similarity than item-item similarity factors. On the three datasets, the best LLM achieves a HitRatio@1 score of 0.5540, 0.3472, and 0.2114 over the item-item factor respectively. But, they are only 0.1914, 0.2966, and 0.1994 over the user-user factor respectively. This might be due to that LLMs can only see a limited context of the users~\cite{wu2024coral}.

\subsection{Mental World of LLMs in RecSys: Ablation}\label{exp:ablation}

\begin{figure}[bt!]
\centering
\begin{subfigure}{.235\textwidth}
\includegraphics[width=0.925\linewidth]{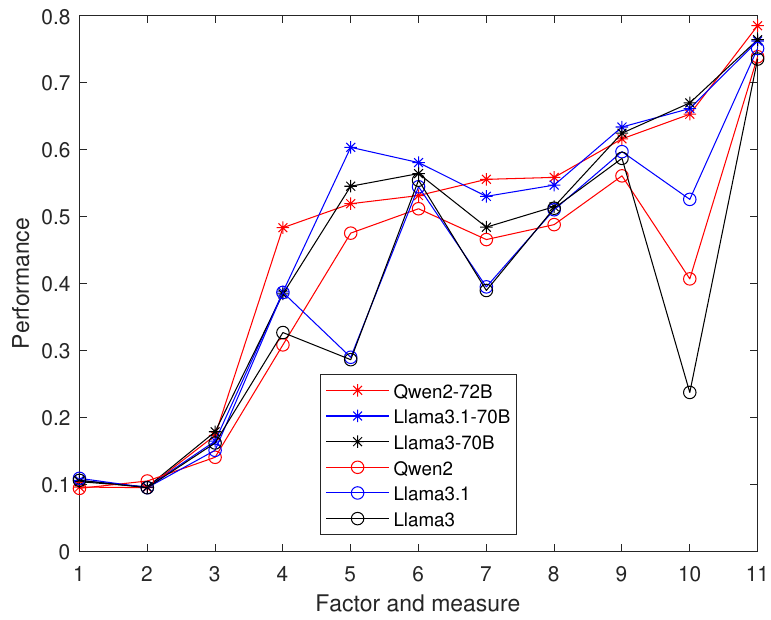}
\end{subfigure}%
\centering
\begin{subfigure}{.25\textwidth}
\includegraphics[width=0.91\linewidth]{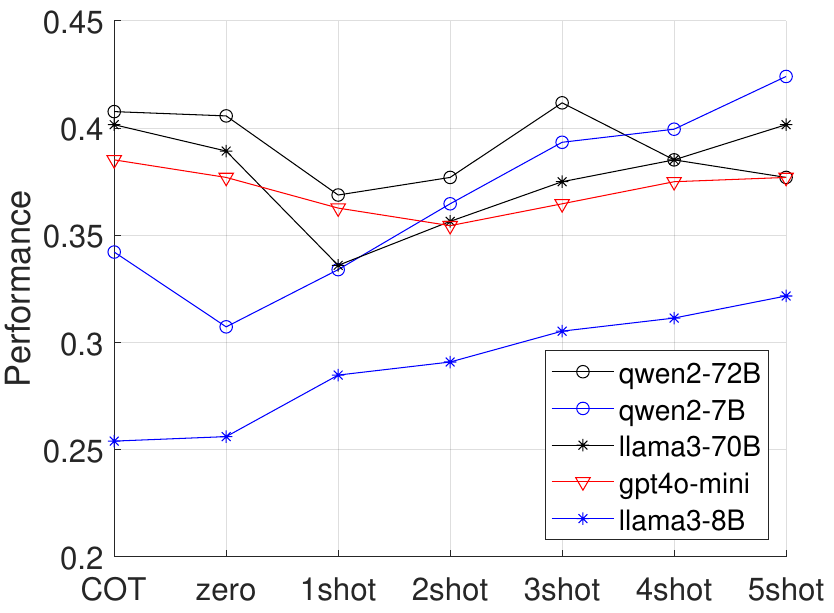}
\end{subfigure}%
\caption{Left: Model Sizes. Right: Prompting Setting. }
\label{fig:largerModels-Fewshot}
\end{figure}

\noindent
{\bf Do larger LLMs show a better mental world in RecSys?} On the largest LRWorld-Amazon dataset, we evaluate on LLMs with an order of magnitude larger, i.e., Qwen2-72B~\cite{qwen2-72b}, Llama3-70B~\cite{llama3-70b} and Llama3.1-70B~\cite{llama3dot1-paper}. Result is shown in Figure~\ref{fig:largerModels-Fewshot}. We can see that larger LLMs do not consistently achieve a significant improvement over their smaller counterparts over the 8 out 11 factors and occasionally degrades (see Table~\ref{tab:larger-model-improvement} for the index of x-axis and detailed values). This might be due to that once LLMs are large enough, the differences in training data and strategies are instead the key role in determining performance on RecSys.

\begin{figure}[bt!]
\begin{subfigure}{.23\textwidth}
  \centering
  \includegraphics[width=0.85\linewidth]{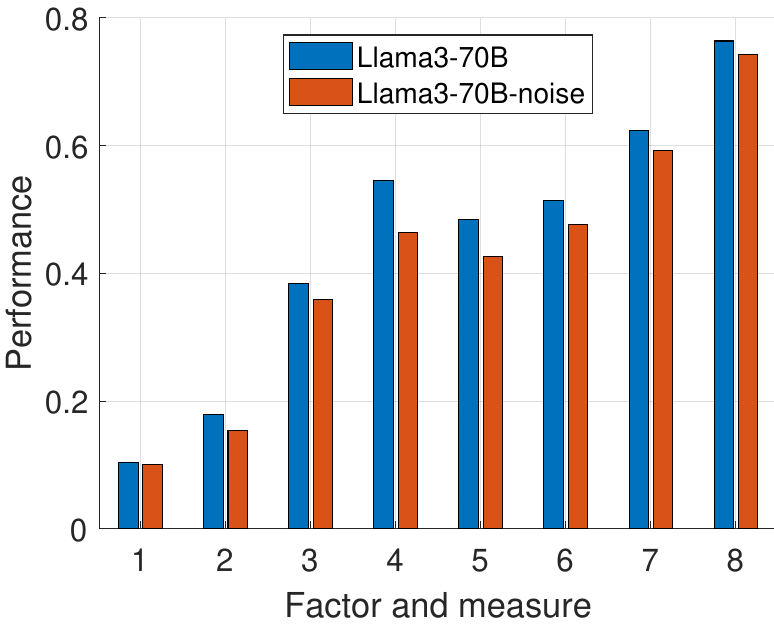}
\end{subfigure}
\begin{subfigure}{.23\textwidth}
  \centering
  \includegraphics[width=0.85\linewidth]{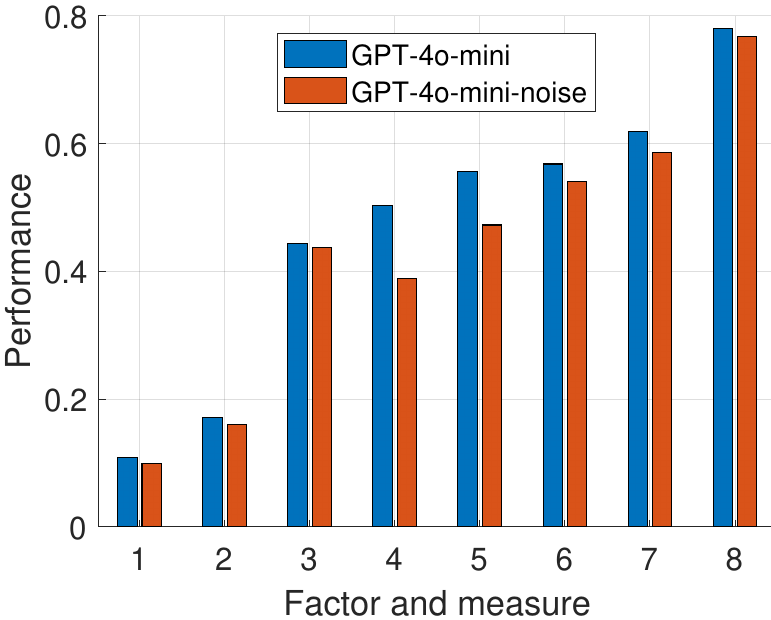}
\end{subfigure}%
\caption{Robustness to Noisy User Profile. (see more in Fig~\ref{fig:noisy-profiles-more})}
\label{fig:noisy-profiles}
\end{figure}

\noindent
{\bf Will advanced promptings influence LLMs in RecSys a lot?} As shown in previous works~\cite{brown2020language,kojima2022large,wei2022chain}, few-shot learning and chain-of-thought (COT) promptings may improve the performance of LLMs. We investigate them in RecSys on the factor of text knowledge reasoning which requires LLMs to infer users' preferences from the movie logline. For the few-shot prompting, we conduct $k$-shot setting and the prompt is listed in Fig~\ref{fig:prompts-fewshot-logline}. For the COT prompting, we append ``Let's think step by step.'' to the zero-shot prompt. The results evaluated on the LRWorld-Netflix dataset are shown in Fig~\ref{fig:largerModels-Fewshot}. We can see that COT benefits qwen2-7B and is almost no effect on other four. As with increasing $k$-shot, i) both llama3-8B and qwen2-7B benefit a lot, ii) llama3-70B first degrades and then improves, iii) qwen2-72B is not stable and peaks at $k=3$, and iv) gpt4o-mini is generally stable. In summary, the smaller LLMs are more likely to benefit from more examples while their larger counterparts are not. This might inspire us to choose higher-relevant few-shot examples for each user. More results on the factor of memory-based user-user similarity retrieval are shown in Fig~\ref{fig:amazon-mem-user-user}.

\noindent
{\bf Are LLMs robust to a noisy world in RecSys?} Take the LRWorld-Amazon dataset for example, there are 8 out of 11 factors that require LLMs to reason based on the user's history. Given a user's historical sequence, say $(i_1,i_2,i_3,i_4,i_5)$, we randomly pick an index (say 3) and replace it with an item (say $i^{-}$) from the item pool to form a noisy and fake profile $(i_1,i_2,i^{-},i_4,i_5)$. We evaluate on both larger and smaller LLMs in this noisy scenario, and results are shown in Figure~\ref{fig:noisy-profiles}. All evaluated LLMs are decreasing in the noisy scenario. Generally, both the larger and smaller LLMs are robust to the noise over the 8 factors except for the factor 4 of multimodal knowledge (i.e. product image) reasoning  (see Table~\ref{tab:noisy-profiles} for the index of x-axis and detailed values). On this factor, Qwen2-72B drops by 21.3\% and Llama3-70B by 17.4\%. Furthermore, the larger LLMs seem more sensitive to the noise than their smaller counterparts, in contrast to Qwen2-7B by 13.9\% and Llama3-8B by 16.0\%. As for GPT-4o-mini, it drops by 29.3\%.

\begin{table}[tb!]
\caption{Effect of Textual and Visual Multi-modality.}
\label{fig:multimodal}
\resizebox{.47\textwidth}{!}{
\begin{tabular}{l|c|c|c}
\toprule
 LlaVA-1.6       & LRWorld-Amazon  & LRWorld-MovieLens & LRWorld-Netflix \\
\midrule
Image's text desc Only     & 0.3442  & 0.3364    & 0.2893  \\ 
Image's text desc + Visual & 0.4577  & 0.4018    & 0.3125  \\ \hline
Improvement                & 32.98\% & 19.44\%   & 8.02\% \\
\bottomrule
\end{tabular}
}

\end{table}

\begin{table}[tb!]
\caption{Effect of Different Versions of OpenAI's GPTs.}
\label{tab:gpts}
\resizebox{.5\textwidth}{!}{
\begin{tabular}{l|c|c|c|c|c|l}
\toprule
Mem item-item & GPT-3.5 & GPT-4o-mini & GPT-4o  & GPT-4-Turbo & GPT-4  & \begin{tabular}[c]{@{}l@{}}Best result of other LLMs \\  (excluding OpenAI's GPTs)\end{tabular} \\
\midrule
LRWorld-Amazon   & 0.4622  & 0.5508      & 0.5836 & 0.5442      & \textbf{0.5901} & 0.5540 (Llama-3)  \\ \hline
LRWorld-Netflix  & 0.1806  & 0.1784      & 0.1828 & 0.1629      & 0.1740 & \textbf{0.2114} (Mistral)   \\         
\bottomrule                                   
\end{tabular}
}
\end{table}

\begin{table}[tb!]
\caption{Variances over Five Random Experiments.}
\label{tab:main-run5-mean-variance}
\resizebox{.44\textwidth}{!}{
\begin{tabular}{l|cc|cc|cc}
\toprule
\multicolumn{1}{l|}{\multirow{2}{*}{Factors}} & \multicolumn{2}{c|}{KG \textless{}movie, cast\textgreater{}} & \multicolumn{2}{c|}{KG \textless{}movie, director\textgreater{}} & \multicolumn{2}{c}{KG \textless{}movie, genre\textgreater{}} \\ \cline{2-7} 
\multicolumn{1}{c|}{}                  & \multicolumn{1}{c|}{mean}            & variance           & \multicolumn{1}{c|}{mean}              & variance             & \multicolumn{1}{c|}{mean}            & variance            \\ \midrule 
Falcon                                  & \multicolumn{1}{l|}{25.11}           & 0.023              & \multicolumn{1}{l|}{24.57}             & 0.030                 & \multicolumn{1}{l|}{25.99}           & 0.074              \\ \hline
Vicuna 1.5                               & \multicolumn{1}{l|}{25.11}           & 0.023              & \multicolumn{1}{l|}{25.18}             & 0.046                & \multicolumn{1}{l|}{25.45}            & 0.027              \\ \hline
FLAN-T5                                  & \multicolumn{1}{l|}{25.04}           & 0.008              & \multicolumn{1}{l|}{50.91}             & 0.043                & \multicolumn{1}{l|}{42.55}            & 0.046              \\ \hline
Llama-3                                  & \multicolumn{1}{l|}{52.59}           & 0.017              & \multicolumn{1}{l|}{56.08}             & 0.114                & \multicolumn{1}{l|}{44.78}            & 0.214              \\ \hline
Llama-3.1                                & \multicolumn{1}{l|}{44.10}            & 0.107              & \multicolumn{1}{l|}{41.75}             & 0.039                & \multicolumn{1}{l|}{44.31}            & 0.091              \\ \hline
Phi-3                                    & \multicolumn{1}{l|}{44.57}           & 0.039              & \multicolumn{1}{l|}{61.55}             & 0.013                & \multicolumn{1}{l|}{51.78}            & 0.006              \\ \hline
Mistral                                 & \multicolumn{1}{l|}{56.63}           & 0.079              & \multicolumn{1}{l|}{68.14}             & 0.022                & \multicolumn{1}{l|}{36.02}            & 0.054              \\ \hline
Qwen-2                                   & \multicolumn{1}{l|}{47.81}           & 0.072              & \multicolumn{1}{l|}{63.22}             & 0.013                & \multicolumn{1}{l|}{47.74}            & 0.051              \\ \hline
LlaVA 1.6                                & \multicolumn{1}{l|}{56.22}           & 0.066              & \multicolumn{1}{l|}{70.37}             & 0.004                & \multicolumn{1}{l|}{41.21}            & 0.055              \\ \hline
GPT-4o-mini                              & \multicolumn{1}{l|}{63.23}           & 0.058              & \multicolumn{1}{l|}{70.77}             & 0.009                & \multicolumn{1}{l|}{50.83}           & 0.091              \\ 
\bottomrule   
\end{tabular}
}
\end{table}

\noindent
{\bf How helpful is multimodality for LLMs in RecSys?} For the factor of multimodal knowledge reasoning, the movie posters and product images are firstly converted into their text descriptions extracted by the large multimodal model LlaVA 1.6. Besides the image's text description, what if the image's visual information is also provided for? We investigate it by feeding the latest item's image of a user's sequence into LlaVA. In other words, see Fig~\ref{fig:poster2desc}, both the text description and the visual image of the {\it item 5} are provided for the LlaVA using the prompt {\it [Knowledgeability] multimodal knowledge} of Fig~\ref{fig:prompts-list}. The result is shown in Table~\ref{fig:multimodal}. We can see that multimodal models are promising in RecSys~\cite{geng2023vip5}.

\noindent
{\bf Do OpenAI's GPT-4s change the game?} On the challenging factor of memory-based item-item similarity retrieval, we investigate whether OpenAI's other three GPT-4s can change the game substantially. The result is shown in Table~\ref{tab:gpts}. We can see that, the other three GPT-4s do not consistently achieve a significant improvement over GPT-4o-mini (and the best open-source/open-weight LLMs). Since LRWorld-MovieLens belongs to the same movie domain with LRWorld-Netflix, we omit its result due to high cost. A case study on their predictions is shown in Table~\ref{tab:gpts-errors} and explanation in Fig~\ref{fig:error-gpts-explanation}. 

\noindent
{\bf Position bias of answers for LLMs in RecSys} We perform five random experiments by inserting the correct answer into a random order of candidate options per example. The result is shown in Table~\ref{tab:main-run5-mean-variance}. We can see the variance across random orders of correct answers is relatively small. All of them are less than 0.1 except for the Llama models. See Appendix~\ref{appendix:position-bias} for in-depth analysis.

\noindent
{\bf More analyses in Appendix} Prediction error distribution and item popularity bias are shown in  Fig~\ref{fig:error-distribution-item-popularity-inter_user2matchedItems}, missed rates in  Table~\ref{tab:Amazon-missed-rate}-\ref{tab:Netflix-missed-rate}, and inference time in Table~\ref{tab:Amazon-time}. Due to page limit, see Appendix for details (\ref{appendix:prompt-template} - \ref{appendix:related-work}).

\subsection{Pros and Cons of Domain-Specific Fine-tuned Recommender LLMs}\label{exp:lrworld-lastfm}

Considering the domain-specific fine-tuning, we investigate on a recent proposed LlaRA model finetuned from the Llama2-7B backbone~\cite{liao2024llara}. It cannot be directly applied to our three LRWorld (Amazon/MovieLens25M/Netflix) datasets since it trains specific token embeddings from different datasets (LastFM, Steam, and MovieLens100K). Therefore, we evaluate on the same testing set of LastFM dataset evaluated by LlaRA. Since there is no user id and only the sequences of listened artists are available, we build a mini LRWorld (LastFM) dataset consisting of two factors. (i) {\bf <history, next-item>} (item=artist): predict the artist to be next listened based on the previous listening history, and (ii) {\bf <entity, attribute>} (entity=artist, attribute=genre): predict the music style of the artist, e.g., metal, rock, pop et al. The <entity, attribute> is equivalent to {\bf <history, genre-of-next-artist>} which is to predict the genre of the next-artist instead of the artist itself. Since the released dataset of LlaRA did not have the genre information\footnote{\url{https://github.com/ljy0ustc/LLaRA/tree/main}}, we get the metadata of artists by aligning with the HetRec 2011 Last.FM dataset\footnote{https://grouplens.org/datasets/hetrec-2011/}. Statistics of final evaluation LastFM dataset are shown in Table~\ref{tab:statistics-lrworld-lastfm}. Note, LlaRA is finetuned on <history, next-artist> task only, and not on <artist, genre>. We reuse the evaluation code released by LlaRA to compute the metrics (HitRatio@1). The result is shown in Table~\ref{tab:domain-specific-llms-recsys}.

\begin{table}[tb!]
\caption{Statistics of the Mini LRWorld (LastFM) Dataset.}
\label{tab:statistics-lrworld-lastfm}
\resizebox{.5\textwidth}{!}{
\begin{tabular}{l|c|c|c|c}
\toprule
\#artist  & \#listenings  & \#genre & \#<history,next-artist>  & \#<history, genre-of-next-artist>  \\ \midrule 
4,606     & 50,800          & 362     & 122   & 122   \\ 
\bottomrule 
\end{tabular}
}
\end{table}

\begin{table}[tb!]
\caption{Pros and Cons of Domain-specific RecSys LLMs.}
\label{tab:domain-specific-llms-recsys}
\resizebox{.5\textwidth}{!}{
\begin{tabular}{l|c|c|c|c|c|c}
\toprule
Factors  & Phi-3  & Llama-3 & Qwen-2  & GPT-4o-mini & GPT-4o & LlaRA  \\ \midrule 
<history, next-artist>     & 0.3114 & 0.2868 & 0.3524 & 0.4590  & \textbf{0.5491}     & 0.5081 \\ \hline
<artist, genre> & 0.4918 & 0.4754 & 0.5082 & 0.5163   & \textbf{0.5492}    & 0.3560  \\ 
\bottomrule 
\end{tabular}
}
\end{table}

On the finetuned task of <history, next-artist>, LlaRA achieves very good performance. It outperforms its backbone Llama family (its backbone is Llama 2 but it outperforms Llama 3 with a large margin). It also outperforms OpenAI's GPT-4o-mini and is comparable with the advanced GPT-4o. This impressive result of LlaRA is achieved with the open-source (open-weight) general LLMs. Introducing domain-specific fine-tuning can enhance the domain knowledge coverage of the general LLMs on the specific recommendation tasks.

On the newly unseen task of <history, genre-of-next-artist>, LlaRA gets stuck. Note that, LlaRA responds a valid response without rejecting the questions and therefore the performance of LlaRA is not caused by the missed rate. LlaRA shows a big drop over the general LLMs like Phi-3 and Qwen-2, in contrast that LlaRA outperforms them on the finetuned task with a large margin. Furthermore, LlaRA gets much worse than its backbone Llama family. This shows that the domain-specific finetuned recommender LLMs may have a high risk of catastrophic forgetting. It reminds us that a proper evaluation of LLMs in RecSys is not trivial and worthy of carefully investigating. Our comprehensive LRWorld benchmark has made a step for this direction: the Personalization scale is not enough, and other key scales (Association and Knowledgeability) are necessary for fully picturing the mental world of LLMs in RecSys.

\section{Related Work}\label{paper:related-work}

There are vibrant studies and surveys on LLMs for RecSys~\cite{wu2023survey,wang2023generative,li2024survey}. We briefly review related works from the perspective of evaluation-oriented LLMs in RecSys, as shown in Table~\ref{tab:related-work-list}. We organize them by the evaluation factors and data fields. 

The evaluation factors of existing works are basically the forms of <user,item,rating> and <user-history,next-item>, corresponding to the classical rating prediction and top-N recommendation tasks. LLMs complete these tasks by prompts of zero-shot, chain-of-thought, and few-shot~\cite{kang2023llms,li2023e4srec,liu2023chatgpt,sanner2023large,harte2023leveraging,dai2023uncovering,di2023evaluating,hou2024large}. Beyond accuracy metrics, the fairness, explanation, summarization, and interactive simulators are also investigated. It needs more data fields of contents, utterances, metadata, and demographics besides the basic (user,item,rating/review/title)~\cite{li2023personalized,lian2024recai,liu2023chatgpt,zhang2023chatgpt,he2023large,wang2023rethinking,ramos2024preference,xu2024fairly}. They have either limited evaluation factors or data fields.

The ProLLM4Rec employed LLMs for RecSys by delving into prompts on task description, user interest modeling, candidate items construction, and prompting strategies~\cite{xu2024prompting}. The LLMRec framework designed to benchmark LLMs on recommendation tasks of rating prediction, sequential recommendation, direct recommendation, explanation generation, and review summarization in understanding the capabilities of LLMs in recommendation scenarios~\cite{liu2023llmrec}. 

Following the direction of evaluation-oriented research, we describe the mental world of LLMs in RecSys beyond the {\it Personalization} scale of the above works.  We further decompose the {\it Personalization} scale into fine-grained factors of shallow (memory-based) vs deep (neural embeddings) and item-centered vs user-centered. We are the first to fully picture the mental world of LLMs in RecSys by proposing the novel scales of {\it Association} and {\it Knowledgeability} (knowledge graph, hierarchical taxonomies, texts, and multimodal knowledge). The proposed LRWorld benchmark has a wide coverage of ten factors and 31 measures (tasks) with 14 fields. Based on the large test samples and tens of millions tokens, we conduct a systematic evaluation on dozens of SOTA LLMs with comprehensive ablation studies.

\section{Conclusion and Future Work}

Towards answering the question of where to go next in the era of large language models for recommender systems and whether the traditional RecSys methods should be replaced by LLMs, we have pictured the mental world of LLMs in RecSys based on the proposed LRWorld benchmark. The good, bad, and competitive aspects of dozens of LLMs in RecSys are revealed in experimental findings and comprehensive ablation studies. To be successful, LLMs for RecSys must fulfill a variety of key scales beyond personalization, including association and knowledgeability. 

It seems promising for LLMs in RecSys by combining the good in Association and Knowledgeability to improve the bad in Personalization. It is worthy of combining robustness to noisy profiles and extraction of multimodal knowledge from posters/images to attack cold-start issues in RecSys.

\begin{acks}
This work is supported by the Fundamental Research Funds for Central Universities (ZYTS25091) and National Natural Science Foundation of China (No. 62306220).
\end{acks}



\bibliographystyle{ACM-Reference-Format}
\bibliography{sample-base}

\newpage

\appendix

\begin{table*}[]
\caption{Illustrations on Texts Knowledge Reasoning about Logline of Movies from Our LRWorld (Netflix) Benchmark. Note, each movie is represented as its logline (brief summary of the movie) only without movie title. (go back to Section~\textcolor{blue}{\ref{paper:knowledge}}) }
\label{tab:illustration-logline}
\resizebox{.97\textwidth}{!}{
\begin{tabular}{l|l}
\toprule
\multicolumn{2}{l}{\textbf{User\_1340132   History of Her Watched Movies (logline only):}}                                                                                                                               \\ \midrule
1  & \begin{tabular}[c]{@{}l@{}}CIA agent  Jack Ryan tries to discover why three missing Russian nuclear scientists are  holed up in the Ukraine, communicating with neo-Nazis.\end{tabular}           \\ \hline
2  & \begin{tabular}[c]{@{}l@{}}A veteran cop is tasked with   drafting and training a special weapons and  tactics team, who soon find   themselves up against an international criminal.\end{tabular} \\ \hline
3  & \begin{tabular}[c]{@{}l@{}}An FBI agent makes it his   mission to put cunning con man Frank Abagnale Jr.   behind bars. But Frank not   only eludes capture, he revels in the pursuit.\end{tabular} \\ \hline
4  & \begin{tabular}[c]{@{}l@{}}New undercover cop partners   Starsky and Hutch must overcome their differences  to solve  an important case   with help from street informant Huggy Bear.\end{tabular}  \\ \hline
5  & \begin{tabular}[c]{@{}l@{}}When Melvin and his friends   gather for dinner, their casual conversation takes   a sharp turn  as they begin   to disclose raw truths about themselves.\end{tabular}    \\ 
\midrule \midrule
\multicolumn{2}{l}{\textbf{Which movie that this user is most likely to watch next?}}                                                                                                                   \\ \midrule
A) & \begin{tabular}[c]{@{}l@{}}U.S.   hockey coach Herb Brooks unites a motley crew of college athletes and   turns  them  into a force to be reckoned with at the 1980 Winter Olympics.\end{tabular}    \\ \hline
B) & \begin{tabular}[c]{@{}l@{}}Grad student Helen Lyle   unintentionally summons the Candyman,   a hook-handed creature made flesh by   other people's belief in him.\end{tabular}                      \\ \hline
C) & \begin{tabular}[c]{@{}l@{}}The Bride has three left on   her rampage list: Budd, Elle Driver and Bill himself.  But when she arrives at   Bill's house, she's in for a surprise.\end{tabular}      \\ \hline
D) & \begin{tabular}[c]{@{}l@{}}Sylvester Stallone shot to   fame as Rocky Balboa, an unknown fighter who's   given a shot at fighting world   champ Apollo Creed as a publicity stunt.\end{tabular}     \\ \midrule
\multicolumn{2}{l}{\textbf{Answer: C}}                  \\ 
\bottomrule     \midrule                                                                                                                                                 
\end{tabular}
}
\end{table*}

\begin{table*}[tb!]
\caption{Illustrations on Retrieving A User's Most Matched Movies According To Distances in Their Embedding Space from Our LRWorld (MovieLens) Benchmark. Each movie is originally represented by its embedding (see Fig~\ref{fig:movielens-visualization-cluster}) to compute the distance with the target user, and here we show their meta information (title, year, and genres) for human reading. (go back to Section~\textcolor{blue}{\ref{paper:prompt-design}}) }
\label{tab:example-user2matchedItems}
\resizebox{.63\textwidth}{!}{
\begin{tabular}{l|l}
\toprule
\multicolumn{2}{l}{\textbf{User\_1714139   History of Her Watched Movies:}}                                                                                                                               \\ \midrule
1  & \begin{tabular}[c]{@{}l@{}}Kill Bill: Vol. 1 (2003), Action \& Adventure.\end{tabular}           \\ \hline
2  & \begin{tabular}[c]{@{}l@{}}Clear and Present Danger (1994),  Action \& Adventure|Dramas.\end{tabular} \\ \hline
3  & \begin{tabular}[c]{@{}l@{}}Indiana Jones and the Last Crusade (1989),  Action \&  Adventure|Children \& Family Movies|Classic Movies.\end{tabular} \\ \hline
4  & \begin{tabular}[c]{@{}l@{}}Hitch (2005),  Comedies|Romantic Movies.\end{tabular}  \\ \hline
5  & \begin{tabular}[c]{@{}l@{}}Coach Carter (2005), Dramas|Sports Movies.\end{tabular}    \\ 
\midrule \midrule
\multicolumn{2}{l}{\textbf{Which movie that this user is most likely to watch next?}}                                                                                                                   \\ \midrule 
1 & \begin{tabular}[c]{@{}l@{}}The Longest Yard (1974), Classic Movies|Comedies|Sports Movies.\end{tabular}    \\ \hline
2 & \begin{tabular}[c]{@{}l@{}}Mystic River (2003), Dramas|Thrillers.\end{tabular}                      \\ \hline
3 & \begin{tabular}[c]{@{}l@{}}American Outlaws (2001),  Action \& Adventure.\end{tabular}      \\ \hline
4 & \begin{tabular}[c]{@{}l@{}}Shabd (2005),  Dramas|International Movies|Romantic Movies.\end{tabular}     \\ \hline
5 & \begin{tabular}[c]{@{}l@{}}Kya Kehna (2000),  Dramas|International Movies|Romantic Movies.\end{tabular}    \\ \hline
6 & \begin{tabular}[c]{@{}l@{}}Insan (2005), Action \& Adventure|Dramas|International Movies.\end{tabular}                      \\ \hline
7 & \begin{tabular}[c]{@{}l@{}}Something's Gotta Give (2003), Comedies|Romantic Movies.\end{tabular}      \\ \hline
8 & \begin{tabular}[c]{@{}l@{}}The Amityville Horror (2005),  Horror Movies.\end{tabular}     \\ \hline
9 & \begin{tabular}[c]{@{}l@{}}Biggie \& Tupac (2002),  Documentaries|Music \& Musicals.\end{tabular}    \\ \hline
10 & \begin{tabular}[c]{@{}l@{}}Tremors 4: The Legend Begins (2004),   Comedies|Horror Movies|Sci-Fi \& Fantasy.\end{tabular}                      \\ \midrule \midrule
\multicolumn{2}{l}{\textbf{Answer: 4}}                  \\ 
\bottomrule                                                                                                                                                     
\end{tabular}
}
\end{table*}

 \begin{figure}[bt!]
  \centering
  \includegraphics[width=0.82\linewidth]{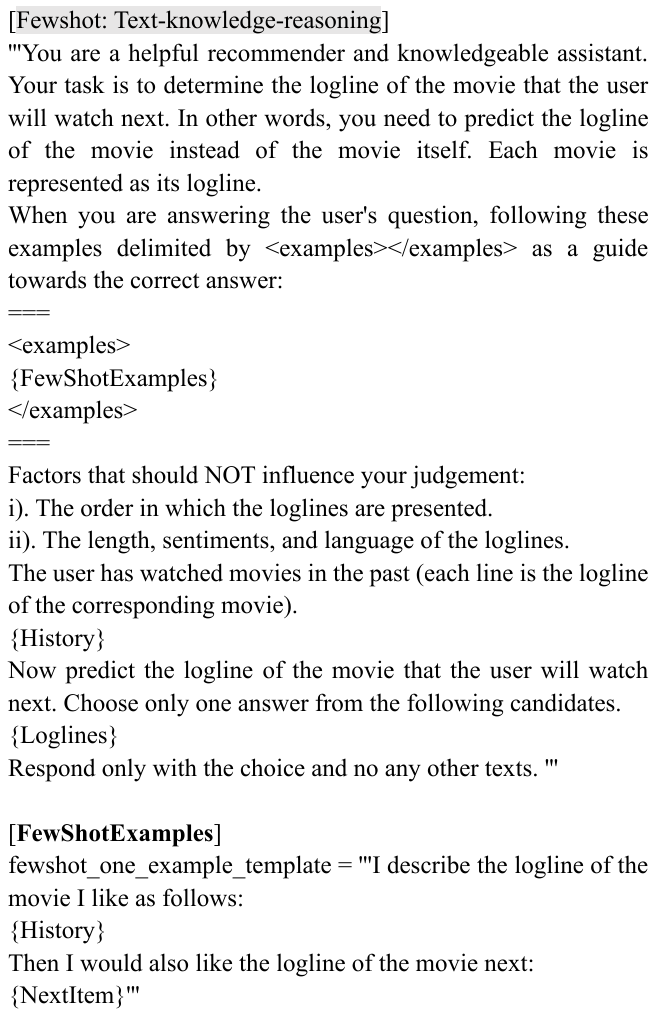}
  \caption{Few-shot Prompts for the Factor of Text Knowledge Reasoning (Movie Logline). See Table~\ref{tab:scales-factors-description} for the detailed description of each scale and factor. (go back to Section~\textcolor{blue}{\ref{exp:ablation}})}
  \label{fig:prompts-fewshot-logline}
\end{figure}

\section{Appendix}\label{appendix}

We report more results and give more detailed analyses. There are tens of subsections (\ref{appendix:prompt-template} - \ref{appendix:related-work}) in this Appendix.

\subsection{Evaluation Promptings Templates}\label{appendix:prompt-template}

\begin{figure*}[tb!]
    \centering
    \begin{subfigure}[t]{0.47\textwidth}
        \centering
        \includegraphics[width=0.93\linewidth]{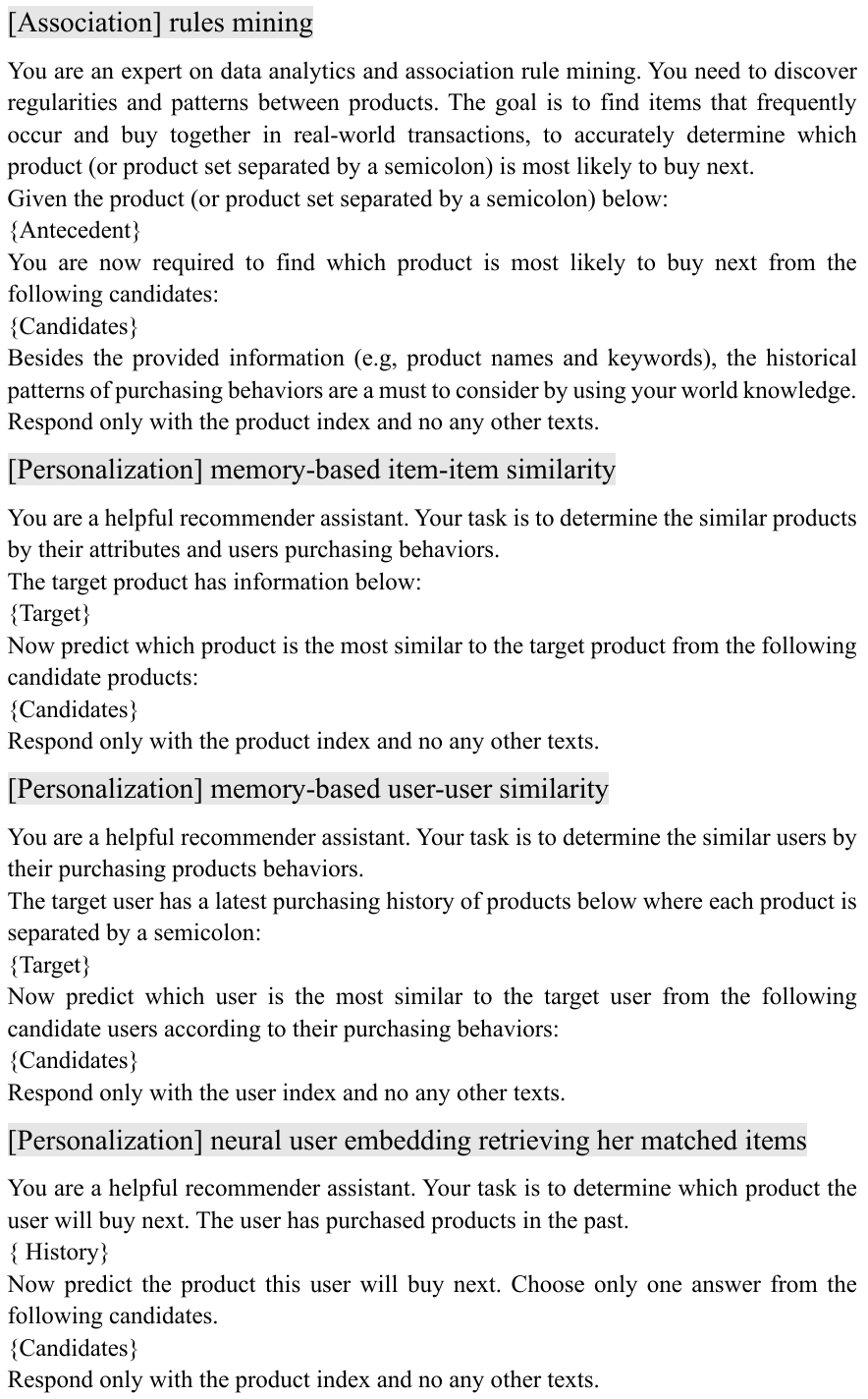}
    \end{subfigure}%
    \begin{subfigure}[t]{0.47\textwidth}
        \centering
        \includegraphics[width=0.93\linewidth]{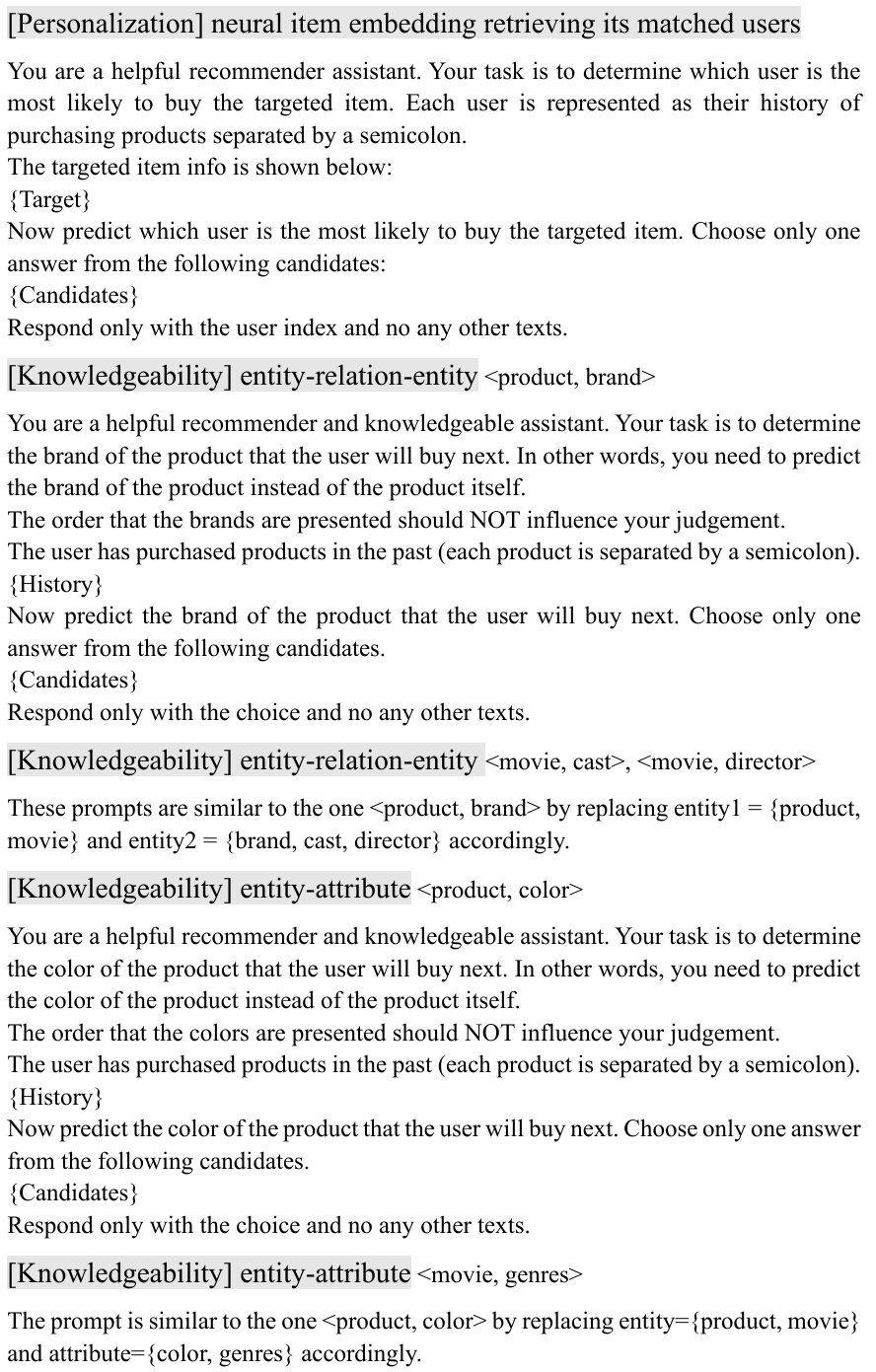}
    \end{subfigure}
    
    \begin{subfigure}[t]{0.47\textwidth}
        \centering
        \includegraphics[width=0.93\linewidth]{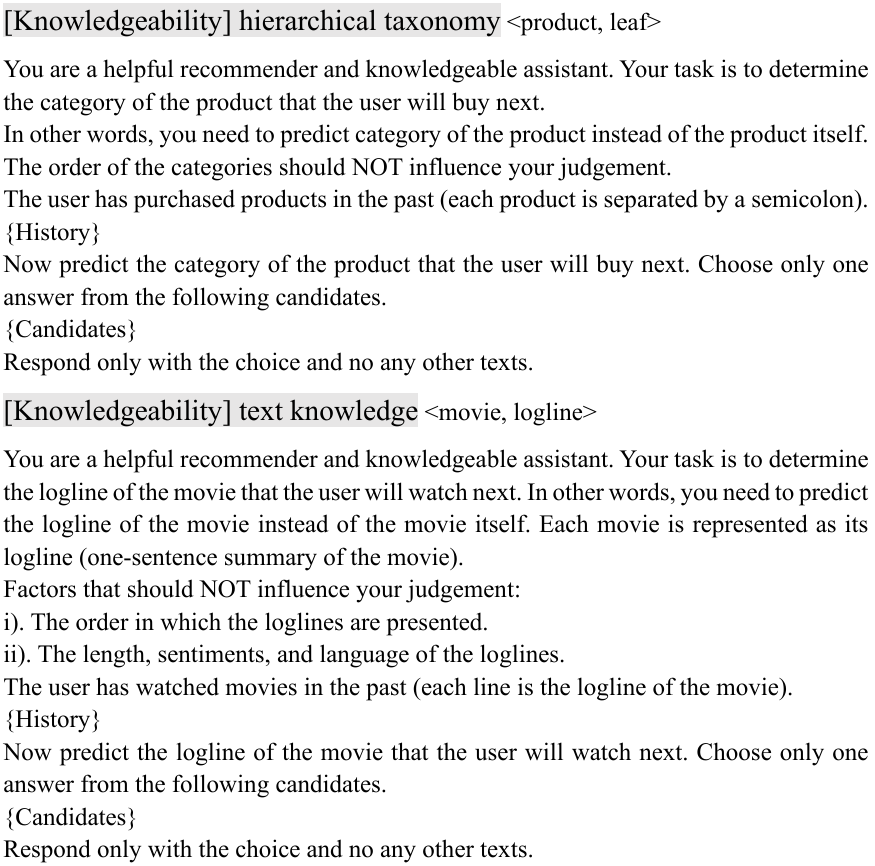}
    \end{subfigure}%
    \begin{subfigure}[t]{0.47\textwidth}
        \centering
        \includegraphics[width=0.93\linewidth]{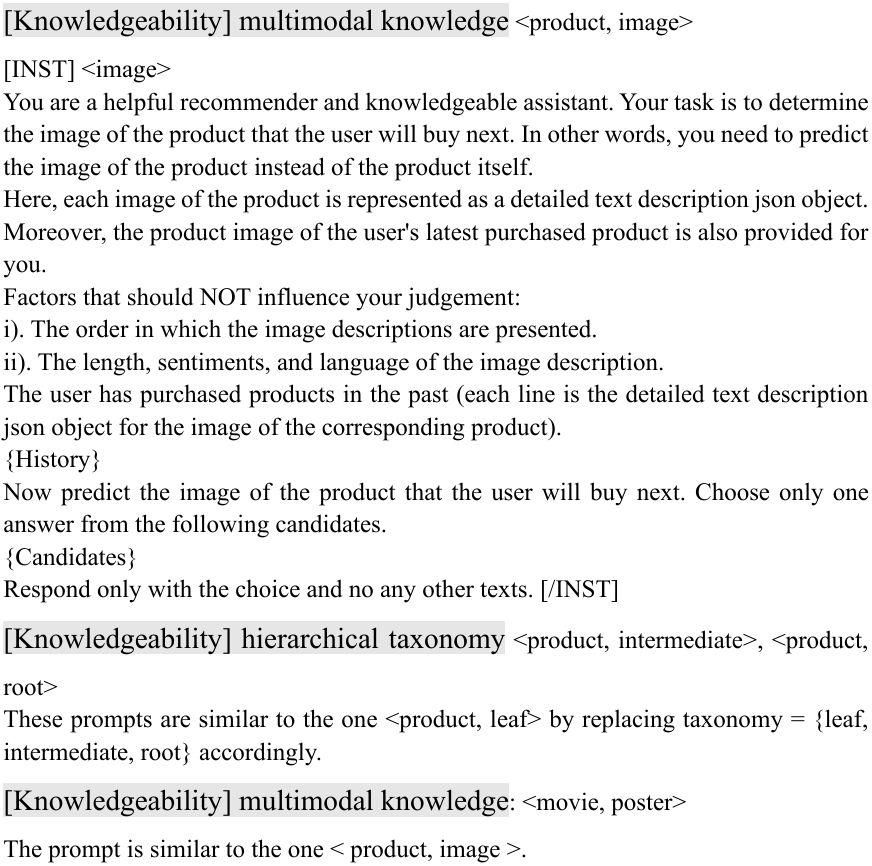}
    \end{subfigure}
    \caption{All of the Prompts Used to Evaluate the Three Main Scales Spanned by Ten Factors with 31 Measures in Total. See Figure~\ref{fig:prompts-fewshot-logline} for few-shot prompts. See Table~\ref{tab:scales-factors-description} for the detailed description of each scale and factor. (go back to Section~\textcolor{blue}{\ref{paper:prompt-design}})} 
\label{fig:prompts-list}
\end{figure*}

All of the evaluation prompting templates are listed in Figure~\ref{fig:prompts-list}. It covers two domains (product domain and movie domain), three datasets (Amazon, MovieLens, and Netflix), three scales (Association, Personalization, and Knowledgeability), 10 factors, 14 fields, and 31 measures (or 31 tasks; each measure corresponding a specific task). 

The chain-of-thought prompts are appending ``Let's think step-by-step'' to the zero-shot evaluation prompts.

The few-shot prompt for the factor of text knowledge (movie's logline) reasoning is shown in Figure~\ref{fig:prompts-fewshot-logline}.

\subsection{More Illustrations and Examples}

Illustration on retrieving a user's most matched items is shown in Table~\ref{tab:example-user2matchedItems}. It belongs to the type of HitRatio@1 question examples. Illustration on texts knowledge reasoning is shown in Table~\ref{tab:illustration-logline}. It belongs to the type of multi-choice question examples.

\begin{table}[tb!]
\caption{Average Accuracy with Variance over Five Random Experiments on the LRWorld (MovieLens) Dataset.  We randomly insert the correct answer into a random order of candidate options for each test example. See Table~\ref{tab:scales-factors-description} for the detailed description of each factor. Scores are scaled by x100 to see the variances more clearly. (go back to Table~\textcolor{blue}{\ref{tab:main-run5-mean-variance}})}
\label{tab:run5-mean-variance}
\resizebox{.45\textwidth}{!}{
\begin{tabular}{l|cc|cc|cc}
\toprule
\multicolumn{1}{l|}{\multirow{2}{*}{Factors}} & \multicolumn{2}{c|}{KG \textless{}movie, cast\textgreater{}} & \multicolumn{2}{c|}{KG \textless{}movie, director\textgreater{}} & \multicolumn{2}{c}{KG \textless{}movie, genre\textgreater{}} \\ \cline{2-7} 
\multicolumn{1}{c|}{}                  & \multicolumn{1}{c|}{mean}            & variance           & \multicolumn{1}{c|}{mean}              & variance             & \multicolumn{1}{c|}{mean}            & variance            \\ \midrule 
Falcon                                  & \multicolumn{1}{l|}{25.11}           & 0.023              & \multicolumn{1}{l|}{24.57}             & 0.030                 & \multicolumn{1}{l|}{25.99}           & 0.074              \\ \hline
Vicuna 1.5                               & \multicolumn{1}{l|}{25.11}           & 0.023              & \multicolumn{1}{l|}{25.18}             & 0.046                & \multicolumn{1}{l|}{25.45}            & 0.027              \\ \hline
FLAN-T5                                  & \multicolumn{1}{l|}{25.04}           & 0.008              & \multicolumn{1}{l|}{50.91}             & 0.043                & \multicolumn{1}{l|}{42.55}            & 0.046              \\ \hline
Llama 3                                  & \multicolumn{1}{l|}{52.59}           & 0.017              & \multicolumn{1}{l|}{56.08}             & 0.114                & \multicolumn{1}{l|}{44.78}            & 0.214              \\ \hline
Llama 3.1                                & \multicolumn{1}{l|}{44.10}            & 0.107              & \multicolumn{1}{l|}{41.75}             & 0.039                & \multicolumn{1}{l|}{44.31}            & 0.091              \\ \hline
Phi-3                                    & \multicolumn{1}{l|}{44.57}           & 0.039              & \multicolumn{1}{l|}{61.55}             & 0.013                & \multicolumn{1}{l|}{51.78}            & 0.006              \\ \hline
Mistral                                 & \multicolumn{1}{l|}{56.63}           & 0.079              & \multicolumn{1}{l|}{68.14}             & 0.022                & \multicolumn{1}{l|}{36.02}            & 0.054              \\ \hline
Qwen-2                                   & \multicolumn{1}{l|}{47.81}           & 0.072              & \multicolumn{1}{l|}{63.22}             & 0.013                & \multicolumn{1}{l|}{47.74}            & 0.051              \\ \hline
LlaVA 1.6                                & \multicolumn{1}{l|}{56.22}           & 0.066              & \multicolumn{1}{l|}{70.37}             & 0.004                & \multicolumn{1}{l|}{41.21}            & 0.055              \\ \hline
GPT-4o-mini                              & \multicolumn{1}{l|}{63.23}           & 0.058              & \multicolumn{1}{l|}{70.77}             & 0.009                & \multicolumn{1}{l|}{50.83}           & 0.091              \\ 
\bottomrule   
\end{tabular}
}
\end{table}

\subsection{Position Bias of LLMs in RecSys}\label{appendix:position-bias}

We analyze the answers' distribution and report multiple random experiments to study the position biases of LLMs in RecSys~\cite{ma2023large}.

\subsubsection{Correct Answer Distributions}

The distributions of the correct answer are shown in the Table~\ref{tab:distribution-Amazon}, Table~\ref{tab:distribution-Movielens}, and Table~\ref{tab:distribution-Netflix}. There are two types of questions (answers). For the Multiple-Choice Question (MCQ) question, the correct answer is randomly inserted into the four candidates. For the HitRate@1 question, the correct answer is randomly inserted into the ten candidates.

\subsubsection{Position Bias of Answers}
To verify if and to what extend the option bias exists on large language models, we perform five random experiments by inserting the correct answer into a random order of candidate options for each test example. The results are shown in Table~\ref{tab:run5-mean-variance} where we report average and the variances. It shows that the variance across different order of correct answers is relatively small. In specific, all of the variances are less than 0.1 except for the Llama models.

\begin{table*}[tb!]
\caption{The Distribution of the Correct Answers on LRWorld (Amazon). There are two types of questions (answers). For the Multiple-Choice Question (MCQ) question, the correct answer and its three negatives are randomly shuffled. For the HitRate@1 question, the correct answer and its nine negatives are randomly shuffled. (go back to Section~\textcolor{blue}{\ref{paper:prompt-design}})}
\label{tab:distribution-Amazon}
\resizebox{.8\textwidth}{!}{
\begin{tabular}{c|c|c|c|c|c|c|c|c|c|c|c}
\toprule
\multicolumn{1}{c|}{Scales} & \multicolumn{1}{c|}{Association} & \multicolumn{4}{c|}{Personalization}                                     & \multicolumn{6}{c}{Knowledgeability}                                                                                    \\ \midrule
Factors & \multicolumn{1}{c|}{\begin{tabular}[c]{@{}c@{}}Assoc \\ rule\end{tabular}} & \multicolumn{1}{c|}{\begin{tabular}[c]{@{}c@{}}Mem \\ item-item\end{tabular}} & \multicolumn{1}{c|}{\begin{tabular}[c]{@{}c@{}}Mem \\ user-user\end{tabular}} & \multicolumn{1}{c|}{\begin{tabular}[c]{@{}c@{}}Neu emb\\ item2users\end{tabular}} & \multicolumn{1}{c|}{\begin{tabular}[c]{@{}c@{}}Neu emb\\ user2items\end{tabular}} & \multicolumn{1}{c|}{\begin{tabular}[c]{@{}c@{}}KG \\ (prod,brand)\end{tabular}} & \multicolumn{1}{c|}{\begin{tabular}[c]{@{}c@{}}KG \\ (prod,color)\end{tabular}} & \multicolumn{1}{c|}{\begin{tabular}[c]{@{}c@{}}Taxo \\ (prod, leaf)\end{tabular}} & \multicolumn{1}{c|}{\begin{tabular}[c]{@{}c@{}}Taxo \\ (prod, interm)\end{tabular}} & \multicolumn{1}{c|}{\begin{tabular}[c]{@{}c@{}}Taxo \\ (prod, root)\end{tabular}} & \multicolumn{1}{c}{\begin{tabular}[c]{@{}c@{}}Multimodal\\ knowledge\end{tabular}} \\
\midrule 
1=A & 10.17\% & 7.87\%  & 10.43\% & 10.30\% & 9.60\%  & 24.93\% & 24.55\% & 23.16\% & 21.24\% & 24.16\% & 23.99\% \\ \hline
2=B & 11.02\% & 9.18\%  & 10.01\% & 9.70\%  & 10.60\% & 26.71\% & 26.40\% & 25.69\% & 27.22\% & 28.22\% & 26.21\% \\ \hline
3=C & 11.02\% & 9.18\%  & 10.63\% & 8.70\%  & 9.70\%  & 24.70\% & 24.82\% & 26.38\% & 27.07\% & 23.70\% & 24.86\% \\ \hline
4=D & 14.41\% & 11.48\% & 9.64\%  & 8.60\%  & 9.10\%  & 23.66\% & 24.24\% & 24.77\% & 24.46\% & 23.93\% & 24.94\% \\ \hline
5   & 7.63\%  & 7.21\%  & 9.98\%  & 11.00\% & 12.20\% &         &         &         &         &         &         \\ 
\cline{1-6}
6   & 11.02\% & 10.82\% & 9.67\%  & 9.50\%  & 10.50\% &         &         &         &         &         &         \\ 
\cline{1-6}
7   & 8.47\%  & 10.49\% & 10.60\% & 10.70\% & 9.60\%  &         &         &         &         &         &         \\
\cline{1-6}
8   & 10.17\% & 13.11\% & 9.45\%  & 9.50\%  & 9.80\%  &         &         &         &         &         &         \\ 
\cline{1-6}
9   & 9.32\%  & 7.87\%  & 9.95\%  & 11.70\% & 9.80\%  &         &         &         &         &         &         \\
\cline{1-6}
10  & 6.78\%  & 12.79\% & 9.62\%  & 10.30\% & 9.10\%  &         &         &         &         &         &         \\ 
\bottomrule                                                                            
\end{tabular}
}
\end{table*}

\begin{table*}[tb!]
\caption{The Distribution of the Correct Answers on LRWorld (MovieLens). }
\label{tab:distribution-Movielens}
\resizebox{.75\textwidth}{!}{
\begin{tabular}{c|c|c|c|c|c|c|c|c|c|c}
\toprule
\multicolumn{1}{c|}{Scales} & \multicolumn{1}{c|}{Association} & \multicolumn{4}{c|}{Personalization}                                     & \multicolumn{5}{c}{Knowledgeability}                                                                                    \\ \midrule 
Factors & \multicolumn{1}{c|}{\begin{tabular}[c]{@{}c@{}}Assoc \\ rule\end{tabular}} & \multicolumn{1}{c|}{\begin{tabular}[c]{@{}c@{}}Mem \\ item-item\end{tabular}} & \multicolumn{1}{c|}{\begin{tabular}[c]{@{}c@{}}Mem \\ user-user\end{tabular}} & \multicolumn{1}{c|}{\begin{tabular}[c]{@{}c@{}}Neu emb\\ item2users\end{tabular}} & \multicolumn{1}{c|}{\begin{tabular}[c]{@{}c@{}}Neu emb\\ user2items\end{tabular}} & \multicolumn{1}{c|}{\begin{tabular}[c]{@{}c@{}}KG \\ (movie,cast)\end{tabular}} & \multicolumn{1}{c|}{\begin{tabular}[c]{@{}c@{}}KG \\ (movie,director)\end{tabular}} & \multicolumn{1}{c|}{\begin{tabular}[c]{@{}c@{}}KG \\ (movie, genre)\end{tabular}} & \multicolumn{1}{c|}{\begin{tabular}[c]{@{}c@{}}Text \\ knowledge \end{tabular}} & \multicolumn{1}{c}{\begin{tabular}[c]{@{}c@{}}Multimodal\\ knowledge\end{tabular}} \\ 
\midrule 
1=A & 10.04\% & 10.43\% & 10.28\% & 9.40\%  & 8.40\%  & 23.91\% & 22.22\% & 24.58\% & 28.96\% & 23.36\% \\ \hline
2=B & 10.47\% & 8.88\%  & 9.78\%  & 10.40\% & 9.60\%  & 29.29\% & 30.98\% & 30.98\% & 25.59\% & 29.91\% \\ \hline
3=C & 9.64\%  & 10.39\% & 10.84\% & 9.80\%  & 10.60\% & 23.91\% & 20.54\% & 20.20\% & 23.23\% & 21.50\% \\ \hline
4=D & 9.51\%  & 10.69\% & 9.83\%  & 9.60\%  & 9.20\%  & 22.90\% & 26.26\% & 24.24\% & 22.22\% & 25.23\% \\ \hline
5   & 10.22\% & 9.95\%  & 10.17\% & 12.10\% & 11.00\% &         &         &         &         &         \\ \cline{1-6}
6   & 9.98\%  & 9.76\%  & 11.06\% & 10.50\% & 11.90\% &         &         &         &         &         \\ \cline{1-6}
7   & 10.65\% & 10.06\% & 9.22\%  & 9.50\%  & 9.50\%  &         &         &         &         &         \\ \cline{1-6}
8   & 9.45\%  & 10.10\% & 9.83\%  & 9.50\%  & 9.70\%  &         &         &         &         &         \\ \cline{1-6}
9   & 10.19\% & 10.32\% & 9.94\%  & 9.90\%  & 8.80\%  &         &         &         &         &         \\ \cline{1-6}
10  & 9.85\%  & 9.43\%  & 9.05\%  & 9.30\%  & 11.30\% &         &         &         &         &         \\ 
\bottomrule                                                                            
\end{tabular}
}
\end{table*}

\begin{table*}[tb!]
\caption{The Distribution of the Correct Answers on LRWorld (Netflix). }
\label{tab:distribution-Netflix}
\resizebox{.75\textwidth}{!}{
\begin{tabular}{c|c|c|c|c|c|c|c|c|c|c}
\toprule
\multicolumn{1}{c|}{Scales} & \multicolumn{1}{c|}{Association} & \multicolumn{4}{c|}{Personalization}                                     & \multicolumn{5}{c}{Knowledgeability}                                                                                    \\ \midrule 
Factors & \multicolumn{1}{c|}{\begin{tabular}[c]{@{}c@{}}Assoc \\ rule\end{tabular}} & \multicolumn{1}{c|}{\begin{tabular}[c]{@{}c@{}}Mem \\ item-item\end{tabular}} & \multicolumn{1}{c|}{\begin{tabular}[c]{@{}c@{}}Mem \\ user-user\end{tabular}} & \multicolumn{1}{c|}{\begin{tabular}[c]{@{}c@{}}Neu emb\\ item2users\end{tabular}} & \multicolumn{1}{c|}{\begin{tabular}[c]{@{}c@{}}Neu emb\\ user2items\end{tabular}} & \multicolumn{1}{c|}{\begin{tabular}[c]{@{}c@{}}KG \\ (movie,cast)\end{tabular}} & \multicolumn{1}{c|}{\begin{tabular}[c]{@{}c@{}}KG \\ (movie,director)\end{tabular}} & \multicolumn{1}{c|}{\begin{tabular}[c]{@{}c@{}}KG \\ (movie, genre)\end{tabular}} & \multicolumn{1}{c|}{\begin{tabular}[c]{@{}c@{}}Text \\ knowledge \end{tabular}} & \multicolumn{1}{c}{\begin{tabular}[c]{@{}c@{}}Multimodal\\ knowledge\end{tabular}} \\ 
\midrule 
1=A & 8.80\%  & 6.39\%  & 10.66\% & 9.68\%  & 7.30\%  & 27.87\% & 25.41\% & 26.64\% & 24.31\% & 27.87\% \\ \hline
2=B & 9.98\%  & 11.01\% & 9.87\%  & 11.17\% & 10.60\% & 24.59\% & 27.46\% & 25.82\% & 26.16\% & 24.59\% \\ \hline
3=C & 11.38\% & 14.10\% & 9.55\%  & 10.00\% & 10.60\% & 22.95\% & 22.34\% & 23.77\% & 25.46\% & 22.95\% \\ \hline
4=D & 10.43\% & 10.13\% & 9.94\%  & 8.40\%  & 9.30\%  & 24.59\% & 24.80\% & 23.77\% & 24.07\% & 24.59\% \\ \hline
5   & 9.59\%  & 11.45\% & 9.51\%  & 12.55\% & 10.80\% &         &         &         &         &         \\ \cline{1-6}
6   & 10.48\% & 7.49\%  & 9.97\%  & 10.21\% & 12.90\% &         &         &         &         &         \\ \cline{1-6}
7   & 10.09\% & 10.13\% & 9.55\%  & 10.11\% & 8.80\%  &         &         &         &         &         \\ \cline{1-6}
8   & 10.65\% & 11.45\% & 10.82\% & 9.26\%  & 9.70\%  &         &         &         &         &         \\ \cline{1-6}
9   & 9.08\%  & 8.15\%  & 9.45\%  & 10.00\% & 9.20\%  &         &         &         &         &         \\ \cline{1-6}
10  & 9.53\%  & 9.69\%  & 10.69\% & 8.62\%  & 10.80\% &         &         &         &         &         \\ 
\bottomrule                                                                            
\end{tabular}
}
\end{table*}

\begin{table*}[tb!]
\caption{Details of the Evaluated Large Language Models.  (go back to Section~\textcolor{blue}{\ref{exp:llm-versions}})}
\label{tab:model-version}
\resizebox{.7\textwidth}{!}{
\begin{tabular}{l|r|l|l|l}
\toprule
\textbf{Name} & \textbf{Size} &\textbf{Version}            & \textbf{Organization}   & \textbf{URL}   \\ 
\midrule 
Falcon	& 7B & falcon-7b-instruct	& 	TII, UAE &	\url{https://huggingface.co/tiiuae/falcon-7b-instruct}\\ \hline
Vicuna-1.5	& 7B & 	vicuna-7b-v1.5	& 	LMSYS Org	 &\url{https://huggingface.co/lmsys/vicuna-7b-v1.5}\\ \hline
FLAN-T5	&  2.8B & 	flan-t5-xl		& 	Google	 & \url{https://huggingface.co/google/flan-t5-xl}\\ \hline
Llama-3		& 8B &  Meta-Llama-3-8B-Instruct	& 	Meta		 &\url{https://huggingface.co/meta-llama/Meta-Llama-3-8B-Instruct}\\ \hline
Llama3-70B		&  70B &  Meta-Llama-3-70B-Instruct	& 	Meta		 &\url{https://huggingface.co/meta-llama/Meta-Llama-3-70B-Instruct}\\ \hline
Llama-3.1		&  8B & Meta-Llama-3.1-8B-Instruct	& 	Meta	 	&\url{https://huggingface.co/meta-llama/Meta-Llama-3.1-8B-Instruct}\\ \hline
Llama3.1-70B		&  70B & Meta-Llama-3.1-70B-Instruct	& 	Meta		&\url{https://huggingface.co/meta-llama/Meta-Llama-3.1-70B-Instruct}\\ \hline
Phi-3		&  3.8B & Phi-3-mini-4k-instruct	& 	Microsoft	 	&\url{https://huggingface.co/microsoft/Phi-3-mini-4k-instruct}\\ \hline
Mistral		& 7B &  Mistral-7B-Instruct-v0.3	& 	Mistral AI	 	&\url{https://huggingface.co/mistralai/Mistral-7B-Instruct-v0.3}\\ \hline
Qwen-2		& 7B &  Qwen2-7B-Instruct		& Qianwen, Alibaba 	 	&\url{https://huggingface.co/Qwen/Qwen2-7B-Instruct}\\ \hline
Qwen2-72B		&  72B & Qwen2-72B-Instruct	& 	Qianwen, Alibaba  &\url{https://huggingface.co/Qwen/Qwen2-72B-Instruct}\\ \hline
LlaVA-1.6		&  7B & llava-v1.6-mistral-7b-hf	& \cite{liu2024llavanext}		&\url{https://huggingface.co/llava-hf/llava-v1.6-mistral-7b-hf}\\ \hline
LLaRA		&  7B & fine-tuned llama-2-7B	&  \cite{liao2024llara}		&\url{https://github.com/ljy0ustc/LLaRA}\\ \hline
GPT-3.5-Turbo		&  N/A & gpt-3.5-turbo-0125		& OpenAI	 &\url{https://platform.openai.com/docs/models/gpt-3-5-turbo}\\ \hline
GPT-4o-mini		&  N/A & gpt-4o-mini-2024-07-18	& 	OpenAI	&\url{https://platform.openai.com/docs/models/gpt-4o-mini}\\ \hline
GPT-4o		&  N/A & gpt-4o-2024-05-13	& 	OpenAI	&\url{https://platform.openai.com/docs/models/gpt-4o}\\ \hline
GPT-4-Turbo		&  N/A & gpt-4-turbo-2024-04-09	& 	OpenAI	&\url{https://platform.openai.com/docs/models/gpt-4-turbo-and-gpt-4}\\ \hline
GPT-4		&  N/A & gpt-4-0613	& 	OpenAI	&\url{https://platform.openai.com/docs/models/gpt-4-turbo-and-gpt-4}\\ 
\bottomrule   
\end{tabular}
}
\end{table*}

\begin{table*}[tb!]
\caption{The Statistics of the Tokens Fed Into LLMs on LRWorld (Amazon) based on GPT-3.5-Turbo APIs.}
\label{tab:tokens-Amazon}
\resizebox{.78\textwidth}{!}{
\begin{tabular}{c|c|c|c|c|c|c|c|c|c|c|c}
\toprule
\multicolumn{1}{c|}{Scales} & \multicolumn{1}{c|}{Association} & \multicolumn{4}{c|}{Personalization}                                     & \multicolumn{6}{c}{Knowledgeability}                                                                                    \\ \midrule
Factors & \multicolumn{1}{c|}{\begin{tabular}[c]{@{}c@{}}Assoc \\ rule\end{tabular}} & \multicolumn{1}{c|}{\begin{tabular}[c]{@{}c@{}}Mem \\ item-item\end{tabular}} & \multicolumn{1}{c|}{\begin{tabular}[c]{@{}c@{}}Mem \\ user-user\end{tabular}} & \multicolumn{1}{c|}{\begin{tabular}[c]{@{}c@{}}Neu emb\\ item2users\end{tabular}} & \multicolumn{1}{c|}{\begin{tabular}[c]{@{}c@{}}Neu emb\\ user2items\end{tabular}} & \multicolumn{1}{c|}{\begin{tabular}[c]{@{}c@{}}KG \\ (prod,brand)\end{tabular}} & \multicolumn{1}{c|}{\begin{tabular}[c]{@{}c@{}}KG \\ (prod,color)\end{tabular}} & \multicolumn{1}{c|}{\begin{tabular}[c]{@{}c@{}}Taxo \\ (prod, leaf)\end{tabular}} & \multicolumn{1}{c|}{\begin{tabular}[c]{@{}c@{}}Taxo \\ (prod, interm)\end{tabular}} & \multicolumn{1}{c|}{\begin{tabular}[c]{@{}c@{}}Taxo \\ (prod, root)\end{tabular}} & \multicolumn{1}{c}{\begin{tabular}[c]{@{}c@{}}Multimodal\\ knowledge\end{tabular}} \\
\midrule 
\#tokens & 71479 & 154703 & 3694227 & 979524 & 570655 & 942859 & 846530 & 466104 & 473104 & 460876 & 2350742 \\ 
\bottomrule                                                                            
\end{tabular}
}
\end{table*}

\begin{table*}[tb!]
\caption{The Statistics of the Tokens Fed Into LLMs on LRWorld (MovieLens and Netflix) based on GPT-3.5-Turbo APIs. }
\label{tab:tokens-Movielens-Netflix}
\resizebox{.76\textwidth}{!}{
\begin{tabular}{c|c|c|c|c|c|c|c|c|c|c}
\toprule
\multicolumn{1}{c|}{Scales} & \multicolumn{1}{c|}{Association} & \multicolumn{4}{c|}{Personalization}                                     & \multicolumn{5}{c}{Knowledgeability}                                                                                    \\ \midrule 
Factors & \multicolumn{1}{c|}{\begin{tabular}[c]{@{}c@{}}Assoc \\ rule\end{tabular}} & \multicolumn{1}{c|}{\begin{tabular}[c]{@{}c@{}}Mem \\ item-item\end{tabular}} & \multicolumn{1}{c|}{\begin{tabular}[c]{@{}c@{}}Mem \\ user-user\end{tabular}} & \multicolumn{1}{c|}{\begin{tabular}[c]{@{}c@{}}Neu emb\\ item2users\end{tabular}} & \multicolumn{1}{c|}{\begin{tabular}[c]{@{}c@{}}Neu emb\\ user2items\end{tabular}} & \multicolumn{1}{c|}{\begin{tabular}[c]{@{}c@{}}KG \\ (movie,cast)\end{tabular}} & \multicolumn{1}{c|}{\begin{tabular}[c]{@{}c@{}}KG \\ (movie,director)\end{tabular}} & \multicolumn{1}{c|}{\begin{tabular}[c]{@{}c@{}}KG \\ (movie, genre)\end{tabular}} & \multicolumn{1}{c|}{\begin{tabular}[c]{@{}c@{}}Text \\ knowledge \end{tabular}} & \multicolumn{1}{c}{\begin{tabular}[c]{@{}c@{}}Multimodal\\ knowledge\end{tabular}} \\ 
\midrule 
MovieLens & 1265565	& 763034	& 1756764& 	985658	& 368123& 	137086	& 88973& 	85503& 	141767	& 161070	 \\ \hline
Netflix & 707219 & 137818 & 3017960 & 887440 & 380389 & 227224 & 142241 & 138000 & 234066 & 640184 \\ 
\bottomrule                                                                            
\end{tabular}
}
\end{table*}

\begin{table*}[tb!]
\caption{Inference Time of LLMs on LRWorld (Amazon). The time format is HH:MM:SS.} 
\label{tab:Amazon-time}
\resizebox{.82\textwidth}{!}{
\begin{tabular}{l|c|c|c|c|c|c|c|c|c|c|c}
\toprule
\multicolumn{1}{l|}{Scales} & \multicolumn{1}{c|}{Association} & \multicolumn{4}{c|}{Personalization}                                     & \multicolumn{6}{c}{Knowledgeability}                                                                                    \\ \midrule
Factors & \multicolumn{1}{c|}{\begin{tabular}[c]{@{}c@{}}Assoc \\ rule\end{tabular}} & \multicolumn{1}{c|}{\begin{tabular}[c]{@{}c@{}}Mem \\ item-item\end{tabular}} & \multicolumn{1}{c|}{\begin{tabular}[c]{@{}c@{}}Mem \\ user-user\end{tabular}} & \multicolumn{1}{c|}{\begin{tabular}[c]{@{}c@{}}Neu emb\\ item2users\end{tabular}} & \multicolumn{1}{c|}{\begin{tabular}[c]{@{}c@{}}Neu emb\\ user2items\end{tabular}} & \multicolumn{1}{c|}{\begin{tabular}[c]{@{}c@{}}KG \\ (prod,brand)\end{tabular}} & \multicolumn{1}{c|}{\begin{tabular}[c]{@{}c@{}}KG \\ (prod,color)\end{tabular}} & \multicolumn{1}{c|}{\begin{tabular}[c]{@{}c@{}}Taxo \\ (prod, leaf)\end{tabular}} & \multicolumn{1}{c|}{\begin{tabular}[c]{@{}c@{}}Taxo \\ (prod, interm)\end{tabular}} & \multicolumn{1}{c|}{\begin{tabular}[c]{@{}c@{}}Taxo \\ (prod, root)\end{tabular}} & \multicolumn{1}{c}{\begin{tabular}[c]{@{}c@{}}Multimodal\\ knowledge\end{tabular}} \\
\midrule 
Falcon      & 0:38 & 1:25  & 24:25   & 6:44  & 6:32  & 10:14   & 11:47   & 1:52  & 5:45  & 5:12  & 16:23 \\ \hline
Vicuna-1.5  & 0:30 & 1:08  & 28:05   & 9:53  & 5:44  & 8:39    & 13:07   & 6:15  & 6:16  & 6:13  & 13:57 \\ \hline
FLAN-T5     & 0:25 & 1:25  & 16:35   & 4:49  & 5:00  & 7:51    & 8:00    & 4:15  & 4:19  & 4:20  & 16:32 \\ \hline
Llama-3     & 0:30 & 1:10  & 20:02   & 5:24  & 3:57  & 8:39    & 8:17    & 4:37  & 4:36  & 4:34  & 14:01 \\ \hline
Llama-3.1   & 0:30 & 1:12  & 33:09   & 10:47 & 7:17  & 15:27   & 14:40   & 7:22  & 7:49  & 4:11  & 14:02 \\ \hline
Phi-3       & 0:42 & 1:35  & 23:02   & 6:22  & 5:17  & 12:41   & 11:55   & 6:46  & 6:49  & 6:52  & 15:05 \\ \hline
Mistral     & 0:36 & 1:55  & 28:33   & 7:49  & 6:07  & 13:51   & 13:28   & 8:05  & 8:07  & 7:58  & 19:15 \\ \hline
Qwen-2      & 0:28 & 1:05  & 32:19   & 4:53  & 3:32  & 7:06    & 6:48    & 3:47  & 3:30  & 3:45  & 12:11 \\ \hline
LlaVA-1.6   & 0:56 & 2:15  & 41:43   & 16:28 & 12:28 & 21:54   & 18:06   & 9:41  & 9:50  & 9:50  & 23:22 \\ \hline
GPT-3.5-Turbo     & 6:12 & 15:48 & 3:12:33 & 49:59 & 43:06 & 1:57:04 & 1:06:00 & 24:40 & 13:57 & 24:36 & 48:55 \\ \hline
GPT-4o-mini & 1:53 & 4:41  & 55:24   & 17:09 & 17:25 & 40:57   & 38:26   & 19:30 & 21:59 & 19:31 & 38:19 \\ 
\bottomrule                                                                            
\end{tabular}
}
\end{table*}

\begin{figure}[bt!]
\centering
\begin{subfigure}[t]{0.33\textwidth}
    \centering
    \includegraphics[width=0.9\linewidth]{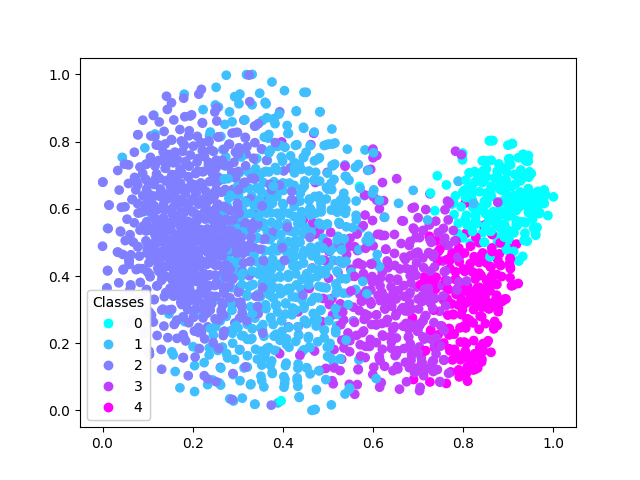}
    \caption{Visualization of Item Embeddings.}
    \label{fig:movielens-visualization-cluster-item}
\end{subfigure}%

\begin{subfigure}[t]{0.33\textwidth}
    \centering
    \includegraphics[width=0.9\linewidth]{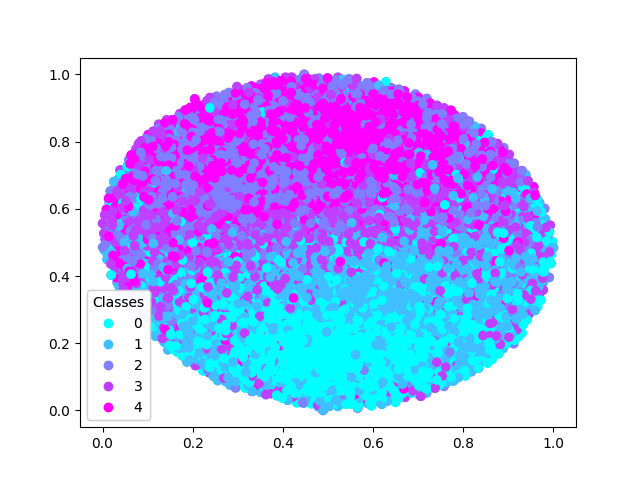}
    \caption{Visualization of User Embeddings.}
    \label{fig:movielens-visualization-cluster-user}
\end{subfigure}
\caption{The t-SNE Visualization of User and Item Embeddings on LRWorld (MovieLens) Dataset. The original dimensionality of the embedding is 64.}
\label{fig:movielens-visualization-cluster}
\end{figure}

\subsection{Details of the Models Evaluated}

The details of the large language models evaluated are shown in the Table~\ref{tab:model-version}. GPT-3.5-turbo and GPT-4o
mini are evaluated via API requests, while other LLMs are evaluated through PyTorch Transformers with Hugging Face. The specific versions listed in the URL are the exactly evaluated ones.

\subsection{Visualization of User and Item Embeddings}

The t-SNE visualization of user and item embeddings on LRWorld (MovieLens) dataset is shown in Figure~\ref{fig:movielens-visualization-cluster}. To get a better visualization effect, the number of clusters is set to 5. In our evaluation experiments, it is set to 10 in order to get more diverse prototypes for testing. These embeddings are learned by neural matrix factorization, a typical recommendation method which can model high-order collaborative relationships between users and items.

\begin{table*}[tb!]
\caption{Missed-rate on LRWorld (Amazon). See Table~\ref{tab:scales-factors-description} for the detailed description of each scale and factor. See Table~\ref{tab:model-version} for the detailed version of the LLMs. The lower the better. (go back to Section~\textcolor{blue}{\ref{exp:ablation}}) } 
\label{tab:Amazon-missed-rate}
\resizebox{.8\textwidth}{!}{
\begin{tabular}{l|c|c|c|c|c|c|c|c|c|c|c}
\toprule
\multicolumn{1}{l|}{Scales} & \multicolumn{1}{c|}{Association} & \multicolumn{4}{c|}{Personalization}                                     & \multicolumn{6}{c}{Knowledgeability}                                                                                    \\ \midrule
Factors & \multicolumn{1}{c|}{\begin{tabular}[c]{@{}c@{}}Assoc \\ rule\end{tabular}} & \multicolumn{1}{c|}{\begin{tabular}[c]{@{}c@{}}Mem \\ item-item\end{tabular}} & \multicolumn{1}{c|}{\begin{tabular}[c]{@{}c@{}}Mem \\ user-user\end{tabular}} & \multicolumn{1}{c|}{\begin{tabular}[c]{@{}c@{}}Neu emb\\ item2users\end{tabular}} & \multicolumn{1}{c|}{\begin{tabular}[c]{@{}c@{}}Neu emb\\ user2items\end{tabular}} & \multicolumn{1}{c|}{\begin{tabular}[c]{@{}c@{}}KG \\ (prod,brand)\end{tabular}} & \multicolumn{1}{c|}{\begin{tabular}[c]{@{}c@{}}KG \\ (prod,color)\end{tabular}} & \multicolumn{1}{c|}{\begin{tabular}[c]{@{}c@{}}Taxo \\ (prod, leaf)\end{tabular}} & \multicolumn{1}{c|}{\begin{tabular}[c]{@{}c@{}}Taxo \\ (prod, interm)\end{tabular}} & \multicolumn{1}{c|}{\begin{tabular}[c]{@{}c@{}}Taxo \\ (prod, root)\end{tabular}} & \multicolumn{1}{c}{\begin{tabular}[c]{@{}c@{}}Multimodal\\ knowledge\end{tabular}} \\
\midrule 
Falcon & 0.0000 & 0.0230 & 0.0000 & 0.0000 & 0.0000 & 0.0266 & 0.0000 & 0.0000 & 0.0000 & 0.0008 & 0.0000  \\ \hline
Vicuna-1.5 & 0.0000 & 0.0000 & 0.0000 & 0.0000 & 0.0000 & 0.0000 & 0.0000 & 0.0000 & 0.0000 & 0.0000 & 0.0000  \\ \hline
FLAN-T5 & 0.0169 & 0.0000 & 0.0011 & 0.0000 & 0.0000 & 0.0313 & 0.0201 & 0.0000 & 0.0000 & 0.0000 & 0.0000  \\ \hline
Llama-3 & 0.0085 & 0.0000 & 0.0000 & 0.0000 & 0.0000 & 0.0000 & 0.0000 & 0.0000 & 0.0000 & 0.0000 & 0.0000  \\ \hline
Llama-3.1 & 0.0000 & 0.0000 & 0.0000 & 0.0000 & 0.0000 & 0.0000 & 0.0000 & 0.0000 & 0.0000 & 0.0000 & 0.0000  \\ \hline
Phi-3 & 0.0000 & 0.0000 & 0.0000 & 0.0020 & 0.0000 & 0.0000 & 0.0077 & 0.0000 & 0.0000 & 0.0000 & 0.0000  \\ \hline
Mistral & 0.0000 & 0.0000 & 0.0000 & 0.0010 & 0.0000 & 0.0000 & 0.0000 & 0.0000 & 0.0000 & 0.0000 & 0.0000  \\ \hline
Qwen-2 & 0.0000 & 0.0000 & 0.0000 & 0.0000 & 0.0000 & 0.0000 & 0.0027 & 0.0107 & 0.0038 & 0.0000 & 0.0115  \\ \hline
LlaVA-1.6 & 0.0000 & 0.0000 & 0.0000 & 0.0000 & 0.0000 & 0.0000 & 0.0000 & 0.0000 & 0.0000 & 0.0000 & 0.0000  \\ \hline
GPT-3.5 & 0.0000 & 0.0000 & 0.0000 & 0.0000 & 0.0210 & 0.0000 & 0.0000 & 0.0000 & 0.0031 & 0.0000 & 0.0000  \\ \hline
GPT-4o-mini & 0.0000 & 0.0000 & 0.0000 & 0.0000 & 0.0000 & 0.0000 & 0.0000 & 0.0000 & 0.0000 & 0.0000 & 0.0000  \\ 
\bottomrule                                                                            
\end{tabular}
}
\end{table*}

\begin{table*}[tb!]
\caption{Missed-rate on LRWorld (MovieLens). See Table~\ref{tab:scales-factors-description} for the detailed description of each scale and factor.}
\label{tab:Movielens-missed-rate}
\resizebox{.78\textwidth}{!}{
\begin{tabular}{l|c|c|c|c|c|c|c|c|c|c}
\toprule
\multicolumn{1}{l|}{Scales} & \multicolumn{1}{c|}{Association} & \multicolumn{4}{c|}{Personalization}                                     & \multicolumn{5}{c}{Knowledgeability}                                                                                    \\ \midrule 
Factors & \multicolumn{1}{c|}{\begin{tabular}[c]{@{}c@{}}Assoc \\ rule\end{tabular}} & \multicolumn{1}{c|}{\begin{tabular}[c]{@{}c@{}}Mem \\ item-item\end{tabular}} & \multicolumn{1}{c|}{\begin{tabular}[c]{@{}c@{}}Mem \\ user-user\end{tabular}} & \multicolumn{1}{c|}{\begin{tabular}[c]{@{}c@{}}Neu emb\\ item2users\end{tabular}} & \multicolumn{1}{c|}{\begin{tabular}[c]{@{}c@{}}Neu emb\\ user2items\end{tabular}} & \multicolumn{1}{c|}{\begin{tabular}[c]{@{}c@{}}KG \\ (movie,cast)\end{tabular}} & \multicolumn{1}{c|}{\begin{tabular}[c]{@{}c@{}}KG \\ (movie,director)\end{tabular}} & \multicolumn{1}{c|}{\begin{tabular}[c]{@{}c@{}}KG \\ (movie, genre)\end{tabular}} & \multicolumn{1}{c|}{\begin{tabular}[c]{@{}c@{}}Text \\ knowledge \end{tabular}} & \multicolumn{1}{c}{\begin{tabular}[c]{@{}c@{}}Multimodal\\ knowledge\end{tabular}} \\ 
\midrule 
Falcon & 0.0000 & 0.0030 & 0.0000 & 0.0000 & 0.0000 & 0.0370 & 0.0000 & 0.0000 & 0.0000 & 0.0000 \\ \hline
Vicuna-1.5 & 0.0006 & 0.0026 & 0.0000 & 0.0000 & 0.0000 & 0.0000 & 0.0000 & 0.0000 & 0.0000 & 0.0000 \\ \hline
FLAN-T5 & 0.0003 & 0.0000 & 0.0006 & 0.0000 & 0.0000 & 0.0034 & 0.0000 & 0.0000 & 0.0000 & 0.0000 \\ \hline
Llama-3 & 0.0482 & 0.0000 & 0.0000 & 0.0000 & 0.0000 & 0.0000 & 0.0000 & 0.0000 & 0.0000 & 0.0000 \\ \hline
Llama-3.1 & 0.0000 & 0.0000 & 0.0000 & 0.0000 & 0.0000 & 0.0000 & 0.0000 & 0.0000 & 0.0000 & 0.0000 \\ \hline
Phi-3 & 0.0000 & 0.0004 & 0.0000 & 0.0000 & 0.0000 & 0.0000 & 0.0000 & 0.0000 & 0.0000 & 0.0000 \\ \hline
Mistral & 0.0003 & 0.0000 & 0.0006 & 0.0000 & 0.0000 & 0.0034 & 0.0000 & 0.0000 & 0.0000 & 0.0000 \\ \hline
Qwen-2 & 0.0000 & 0.0000 & 0.0000 & 0.0000 & 0.0050 & 0.0000 & 0.0000 & 0.0000 & 0.0034 & 0.0000 \\ \hline
LlaVA-1.6 & 0.0377 & 0.0000 & 0.0000 & 0.0000 & 0.0000 & 0.0000 & 0.0000 & 0.0000 & 0.0000 & 0.0000 \\ \hline
GPT-3.5 & 0.0000 & 0.0011 & 0.0000 & 0.0000 & 0.0040 & 0.0000 & 0.0000 & 0.0000 & 0.0000 & 0.0000 \\ \hline
GPT-4o-mini & 0.0000 & 0.0033 & 0.0000 & 0.0000 & 0.0000 & 0.0000 & 0.0000 & 0.0000 & 0.0000 & 0.0000 \\ 
\bottomrule                                                                            
\end{tabular}
}
\end{table*}

\begin{table*}[tb!]
\caption{Missed-rate on LRWorld (Netflix). See Table~\ref{tab:scales-factors-description} for the detailed description of each scale and factor. }
\label{tab:Netflix-missed-rate}
\resizebox{.78\textwidth}{!}{
\begin{tabular}{l|c|c|c|c|c|c|c|c|c|c}
\toprule
\multicolumn{1}{l|}{Scales} & \multicolumn{1}{c|}{Association} & \multicolumn{4}{c|}{Personalization}                                     & \multicolumn{5}{c}{Knowledgeability}                                                                                    \\ \midrule
Factors & \multicolumn{1}{c|}{\begin{tabular}[c]{@{}c@{}}Assoc \\ rule\end{tabular}} & \multicolumn{1}{c|}{\begin{tabular}[c]{@{}c@{}}Mem \\ item-item\end{tabular}} & \multicolumn{1}{c|}{\begin{tabular}[c]{@{}c@{}}Mem \\ user-user\end{tabular}} & \multicolumn{1}{c|}{\begin{tabular}[c]{@{}c@{}}Neu emb\\ item2users\end{tabular}} & \multicolumn{1}{c|}{\begin{tabular}[c]{@{}c@{}}Neu emb\\ user2items\end{tabular}} & \multicolumn{1}{c|}{\begin{tabular}[c]{@{}c@{}}KG \\ (movie,cast)\end{tabular}} & \multicolumn{1}{c|}{\begin{tabular}[c]{@{}c@{}}KG \\ (movie,director)\end{tabular}} & \multicolumn{1}{c|}{\begin{tabular}[c]{@{}c@{}}KG \\ (movie, genre)\end{tabular}} & \multicolumn{1}{c|}{\begin{tabular}[c]{@{}c@{}}Text \\ knowledge \end{tabular}} & \multicolumn{1}{c}{\begin{tabular}[c]{@{}c@{}}Multimodal\\ knowledge\end{tabular}} \\
\midrule 
Falcon & 0.0000 & 0.0110 & 0.0000 & 0.0043 & 0.0000 & 0.0082 & 0.0000 & 0.0000 & 0.0000 & 0.0000 \\ \hline
Vicuna-1.5 & 0.0000 & 0.0000 & 0.0000 & 0.0000 & 0.0000 & 0.0000 & 0.0000 & 0.0000 & 0.0000 & 0.0000 \\ \hline
FLAN-T5 & 0.0006 & 0.0000 & 0.0000 & 0.0000 & 0.0000 & 0.0000 & 0.0041 & 0.0000 & 0.0000 & 0.0046 \\ \hline
Llama-3 & 0.0280 & 0.0000 & 0.0000 & 0.0000 & 0.0000 & 0.0000 & 0.0000 & 0.0000 & 0.0000 & 0.0000 \\ \hline
Llama-3.1 & 0.0000 & 0.0000 & 0.0000 & 0.0000 & 0.0000 & 0.0000 & 0.0000 & 0.0000 & 0.0000 & 0.0116 \\ \hline
Phi-3 & 0.0000 & 0.0000 & 0.0000 & 0.0000 & 0.0000 & 0.0000 & 0.0000 & 0.0000 & 0.0000 & 0.0000 \\ \hline
Mistral & 0.0000 & 0.0000 & 0.0000 & 0.0489 & 0.0000 & 0.0000 & 0.0000 & 0.0000 & 0.0000 & 0.0000 \\ \hline
Qwen-2 & 0.0000 & 0.0000 & 0.0000 & 0.0000 & 0.0030 & 0.0000 & 0.0000 & 0.0000 & 0.0020 & 0.0000 \\ \hline
LlaVA-1.6 & 0.0000 & 0.0000 & 0.0000 & 0.0000 & 0.0000 & 0.0000 & 0.0000 & 0.0000 & 0.0000 & 0.0000 \\ \hline
GPT-3.5 & 0.0000 & 0.0000 & 0.0000 & 0.0000 & 0.0270 & 0.0000 & 0.0000 & 0.0000 & 0.0000 & 0.0000 \\ \hline
GPT-4o-mini & 0.0000 & 0.0000 & 0.0000 & 0.0000 & 0.0000 & 0.0000 & 0.0000 & 0.0000 & 0.0000 & 0.0000 \\ 
\bottomrule                                                                            
\end{tabular}
}
\end{table*}

\subsection{Breakdown of Tokens Statistics}

The statistics of the tokens fed into the LLMs (based on the statistics returned by GPT-3.5-Turbo APIs) for each factor and measure are shown in the Table~\ref{tab:tokens-Amazon} and Table~\ref{tab:tokens-Movielens-Netflix}. These numbers of tokens can be used to compute the cost when calling commercial LLMs' APIs. We have called OpenAI's APIs about 111K times for experimental results and ablation studies.

\subsection{Statistics of Missed Rate of Responses}\label{appendix:miss-rate}

We have explicitly instructed the LLMs to respond with a single choice to compute the accuracy and HitRatio@1 metrics. However, we still observed a very small proportion of responses which are invalid. We implemented a regular regression to search for the valid choice from LLMs' responses. The statistics of these missed rates of LLMs responses are shown in the Table~\ref{tab:Amazon-missed-rate}-\ref{tab:Netflix-missed-rate}. We can see that, most of the responses are valid and only a very small proportion of them have missed. The missed rate is within 5\%.

\begin{figure}[bt!]
  \centering
  \includegraphics[width=0.65\linewidth]{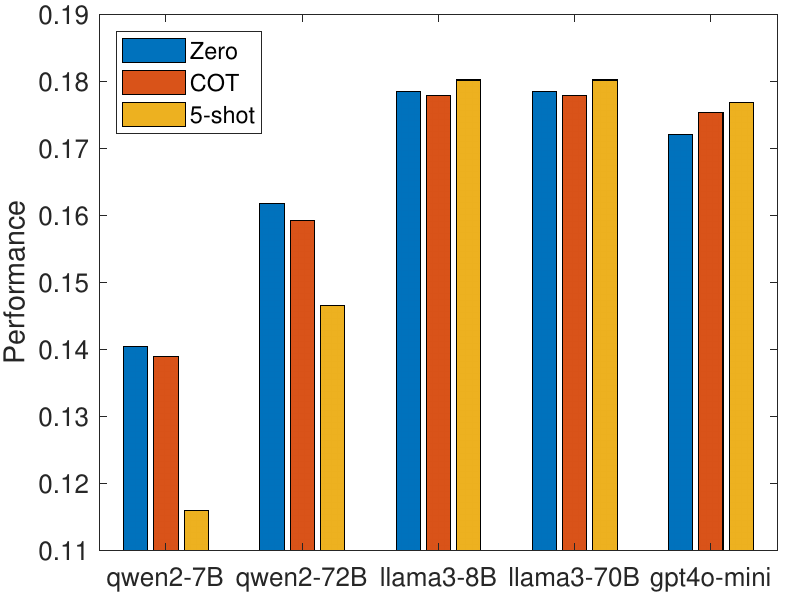}
  \caption{Effect of Prompting Setting, Chain-of-Thought, Zero-shot, and Few-shot. On the factor of memory-based user-user similarity retrieval evaluated on the LRWorld-Netflix dataset. See Table~\ref{tab:scales-factors-description} for the detailed description of each scale and factor. (go back to Figure~\textcolor{blue}{\ref{fig:largerModels-Fewshot}})}
  \label{fig:amazon-mem-user-user}
\end{figure}

\subsection{More Results on Promptings Strategies}

For the effect of advanced promptings, the results on the factor of memory-based user-user similarity retrieval is shown in Figure~\ref{fig:amazon-mem-user-user}. The larger LLMs do not benefit much from few-shot examples.

\subsection{Error Distributions and Popularity Bias}

\begin{figure}[bt!]
\centering
\begin{subfigure}[t]{0.24\textwidth}
    \centering
    \includegraphics[width=0.9\linewidth]{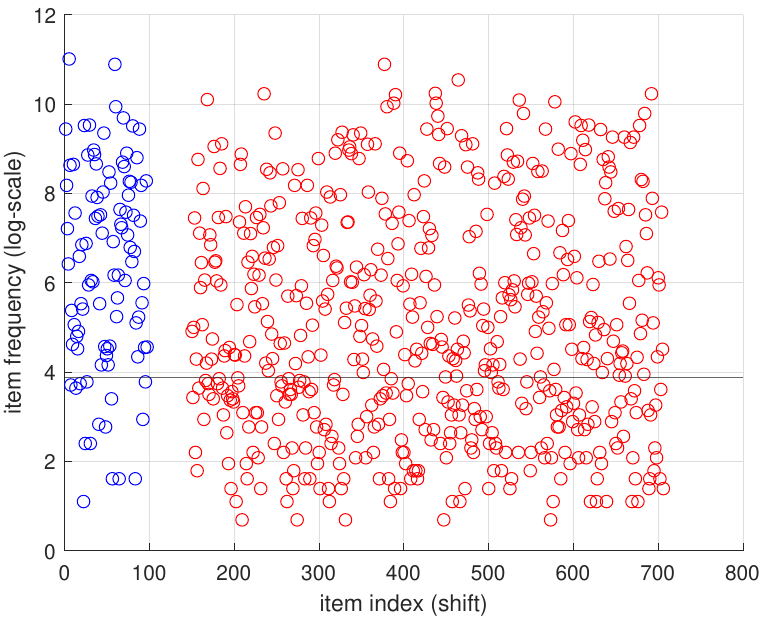}
    \caption{On LRWorld (MovieLens).}
    \label{fig:movielens-inter_user2matchedItems-error}
\end{subfigure}%
\begin{subfigure}[t]{0.24\textwidth}
    \centering
    \includegraphics[width=0.9\linewidth]{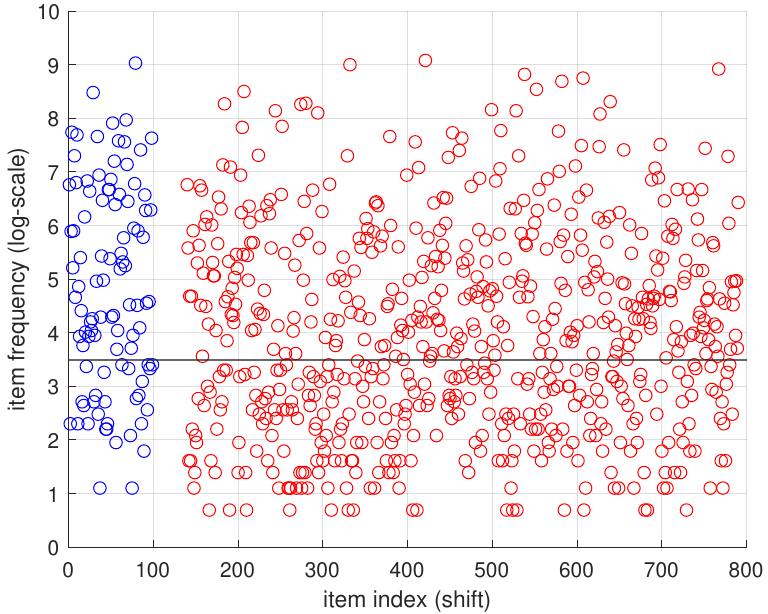}
    \caption{On LRWorld (Amazon).}
    \label{fig:amazon-inter_user2matchedItems-error}
\end{subfigure}
\caption{The Error and Correct Distributions with Their Corresponding Items' Popularity. The \textcolor{red}{error} predictions are marked by red color while the \textcolor{blue}{corrects}  are by blue. The constant y-value (the black horizontal line) is the mean frequency of all items (log-scaled). The x-axis is the item indices which are shifted to separate the error and correct predictions for better visualization. (go back to Section~\textcolor{blue}{\ref{exp:ablation}})}
\label{fig:error-distribution-item-popularity-inter_user2matchedItems}
\end{figure}

Popular items are usually recommended more frequently~\cite{chen2023bias}. Here we analyze the popularity bias of LLMs in RecSys from the perspective of their error distributions correlated with the item popularity degrees. The results are shown in Figure~\ref{fig:error-distribution-item-popularity-inter_user2matchedItems}. The item frequency is a proxy for the popularity of items. We divide the items into two groups, POP-GROUP and NON-POP-GROUP. For items in NON-POP-GROUP, their frequencies are below the mean frequency. And the remaining items belong to POP-GROUP. We now calculate the proportions of NON-POP-GROUP over the error predictions and the correction predictions respectively. Through analyzing their different proportions, we can get a sense of LLMs' popularity bias in RecSys, or their challenges in accurately recommending cold-items.

On the LRWorld (MovieLens) dataset, the mean frequency (log-scaled) is 3.87. For the error predictions, the proportions of items in NON-POP-GROUP are 40.0\%. For the correct predictions, the proportions of items in NON-POP-GROUP are only 15.4\%. In order to maximize the accuracy of recommendations in real-world RecSys, they will recommend more popular items rather than non-popular items. This leads to an issue of bias amplification loop since more and more popular items are recommended while fewer and fewer non-popular items get recommended. 

On the LRWorld (Amazon) dataset, the mean frequency (log-scaled) is 3.49. For the error predictions, the proportions of items in NON-POP-GROUP are 44.5\%. For the correct predictions, the proportions of items in NON-POP-GROUP are 38.6\%. In summary, we can see that this popularity bias still resides in the mental world of LLMs in RecSys.

\begin{table*}[tb!]
\caption{Results on LRWorld (Netflix). See Table~\ref{tab:scales-factors-description} for the detailed description of each scale and factor. The higher the better.  (go back to Table~\textcolor{blue}{\ref{tab:Movielens}})}
\label{tab:Netflix}
\resizebox{.75\textwidth}{!}{
\begin{tabular}{l|c|c|c|c|c|c|c|c|c|c}
\toprule
\multicolumn{1}{l|}{Scales} & \multicolumn{1}{c|}{Association} & \multicolumn{4}{c|}{Personalization}                                     & \multicolumn{5}{c}{Knowledgeability}                                                                                    \\ \midrule
Factors & \multicolumn{1}{c|}{\begin{tabular}[c]{@{}c@{}}Assoc \\ rule\end{tabular}} & \multicolumn{1}{c|}{\begin{tabular}[c]{@{}c@{}}Mem \\ item-item\end{tabular}} & \multicolumn{1}{c|}{\begin{tabular}[c]{@{}c@{}}Mem \\ user-user\end{tabular}} & \multicolumn{1}{c|}{\begin{tabular}[c]{@{}c@{}}Neu emb\\ item2users\end{tabular}} & \multicolumn{1}{c|}{\begin{tabular}[c]{@{}c@{}}Neu emb\\ user2items\end{tabular}} & \multicolumn{1}{c|}{\begin{tabular}[c]{@{}c@{}}KG \\ (movie,cast)\end{tabular}} & \multicolumn{1}{c|}{\begin{tabular}[c]{@{}c@{}}KG \\ (movie,director)\end{tabular}} & \multicolumn{1}{c|}{\begin{tabular}[c]{@{}c@{}}KG \\ (movie, genre)\end{tabular}} & \multicolumn{1}{c|}{\begin{tabular}[c]{@{}c@{}}Text \\ knowledge \end{tabular}} & \multicolumn{1}{c}{\begin{tabular}[c]{@{}c@{}}Multimodal\\ knowledge\end{tabular}} \\
\midrule 
RANDOM & 0.1000 & 0.1000 & 0.1000 & 0.1000 & 0.1000 & 0.2500 & 0.2500 & 0.2500 & 0.2500 & 0.2500\\ \hline
Falcon & 0.0958 & 0.0947 & 0.1075 & 0.0968 & 0.0730 & 0.2397 & 0.2786 & 0.2684 & 0.2663 & 0.2453\\ \hline
Vicuna-1.5 & 0.1597 & 0.1475 & 0.1255 & {\bf 0.1095 } & 0.0730 & 0.2786 & 0.2540 & 0.2725 & 0.2786 & 0.2430\\ \hline
FLAN-T5 & 0.1008 & 0.1651 & 0.1431 & 0.0989 & 0.1200 & 0.3401 & 0.3811 & 0.3176 & 0.3032 & 0.2569\\ \hline
Llama-3 & 0.2303 & 0.2070 & 0.1627 & 0.0882 & 0.1070 & 0.3770 & 0.3934 & 0.3155 & 0.2561 & 0.2962\\ \hline
Llama-3.1 & 0.2937	& 0.1497& 	0.1745& 	0.1021& 	0.1040 & 	0.4098& 	0.3729& 	0.3381& 	0.2561& 	0.2569 \\ \hline
Phi-3 & 0.2158 & 0.1563 & 0.1457 & 0.1074 & {\bf 0.1230 } & 0.3504 & 0.4323 & 0.3258 & 0.2684 & {\bf 0.3425 }\\ \hline
Mistral & 0.2539 & {\bf 0.2114 } & 0.1726 & 0.0797 & 0.1020 & 0.3790 & 0.4733 & 0.3360 & 0.2827 & 0.3125\\ \hline
Qwen-2 & {\bf 0.3206 } & 0.1585 & 0.1706 & 0.0946 & 0.0950 & 0.3647 & 0.4364 & 0.3668 & 0.3073 & 0.2939\\ \hline
LlaVA-1.6 & 0.2264 & 0.1938 & 0.1765 & 0.0968 & 0.1020 & 0.3483 & {\bf 0.4836 } & 0.3360 & 0.3176 & 0.2893\\ \hline
GPT-3.5-Turbo & 0.2494 & 0.1806 & 0.1918 & 0.1000 & 0.0880 & 0.4282 & 0.4405 & 0.3422 & 0.2868 & 0.2777\\ \hline
GPT-4o-mini & 0.3032 & 0.1784 & {\bf 0.1994 } & 0.0914 & 0.1120 & {\bf 0.4631 } & 0.4672 & {\bf 0.3750 } & {\bf 0.3770 } & 0.3032\\ 
\bottomrule                                                                            
\end{tabular}
}
\end{table*}

\begin{table*}[tb!]
\caption{Robustness to Noisy and Fake Profiles of LLMs on LRWorld (Amazon). The factor index is corresponding to the x-axis in Figure~\ref{fig:noisy-profiles}. See Table~\ref{tab:scales-factors-description} for the detailed description of each factor. See Table~\ref{tab:model-version} for the detailed versions of the LLMs. (go back to Figure~\textcolor{blue}{\ref{fig:noisy-profiles}})}
\label{tab:noisy-profiles}
\resizebox{.75\textwidth}{!}{
\begin{tabular}{l|c|c|c|c|c|c|c|c}
\toprule
Factor   Index    & 1                                                                 & 2                                                             & 3                                                                                       & 4                                                                   & 5                                                                                      & 6                                                                                       & 7                                                                                         & 8                                                                                      \\
\midrule
Factor            & \begin{tabular}[c]{@{}c@{}}Neu emb\\      user2items\end{tabular} & \begin{tabular}[c]{@{}c@{}}Mem \\      user-user\end{tabular} & \begin{tabular}[c]{@{}c@{}}Taxo\\      \textless{}prod, root\textgreater{}\end{tabular} & \begin{tabular}[c]{@{}c@{}}Multimodal\\      Knowledge\end{tabular} & \begin{tabular}[c]{@{}c@{}}KG\\      \textless{}prod, color\textgreater{}\end{tabular} & \begin{tabular}[c]{@{}c@{}}Taxo\\      \textless{}prod, leaf\textgreater{}\end{tabular} & \begin{tabular}[c]{@{}c@{}}Taxo\\      \textless{}prod, interm\textgreater{}\end{tabular} & \begin{tabular}[c]{@{}c@{}}KG\\      \textless{}prod, brand\textgreater{}\end{tabular} \\
\midrule 
Llama3-8B         & 0.1060                                                             & 0.1618                                                        & 0.3266                                                                                  & 0.2861                                                              & 0.3894                                                                                 & 0.5122                                                                                  & 0.5866                                                                                    & 0.7348                                                                                 \\
Llama3-8B-noise   & 0.0990                                                              & 0.1381                                                        & 0.3159                                                                                  & 0.2466                                                              & 0.3539                                                                                 & 0.4800                                                                                     & 0.5613                                                                                    & 0.7252                                                                                 \\ \hline
Llama3-70B        & 0.1040                                                              & 0.1785                                                        & 0.3842                                                                                  & 0.5450                                                                & 0.4839                                                                                 & 0.5145                                                                                  & 0.6242                                                                                    & 0.7637                                                                                 \\
Llama3-70B-noise  & 0.1010                                                              & 0.1539                                                        & 0.3588                                                                                  & 0.4640                                                                & 0.4268                                                                                 & 0.4769                                                                                  & 0.5920                                                                                      & 0.7429                                                                                 \\ \hline
Qwen2-7B          & 0.0940                                                              & 0.1404                                                        & 0.3082                                                                                  & 0.475                                                               & 0.4654                                                                                 & 0.4877                                                                                  & 0.5605                                                                                    & 0.7383                                                                                 \\
Qwen2-7B-noise    & 0.0910                                                              & 0.1167                                                        & 0.2852                                                                                  & 0.4169                                                              & 0.4168                                                                                 & 0.4593                                                                                  & 0.5299                                                                                    & 0.7147                                                                                 \\ \hline
Qwen2-72B         & 0.0960                                                              & 0.1731                                                        & 0.4831                                                                                  & 0.5189                                                              & 0.5553                                                                                 & 0.5582                                                                                  & 0.6157                                                                                    & 0.7846                                                                                 \\
Qwen2-72B-noise   & 0.0920                                                              & 0.1520                                                          & 0.4808                                                                                  & 0.4276                                                              & 0.4920                                                                                   & 0.5230                                                                                    & 0.5935                                                                                    & 0.7757                                                                                 \\ \hline
GPT-4o-mini       & 0.1080                                                              & 0.1720                                                          & 0.4432                                                                                  & 0.5031                                                              & 0.5569                                                                                 & 0.5682                                                                                  & 0.6196                                                                                    & 0.7803                                                                                 \\
GPT-4o-mini-noise & 0.0990                                                              & 0.1610                                                          & 0.4371                                                                                  & 0.3889                                                              & 0.4727                                                                                 & 0.5414                                                                                  & 0.5858                                                                                    & 0.7676                \\                                                                
\bottomrule    
\end{tabular}
}
\end{table*}

\subsection{Detailed Versions of the LLMs}\label{appendix:llm-versions}

To fully picture the mental world of LLMs in RecSys, we evaluate on dozens of state-of-the-art LLMs. See Table~\ref{tab:model-version} for details

\begin{itemize}
  \item {\bf GPTs}: OpenAI's {\bf GPT-3.5-turbo} and {\bf GPT-4o mini}~\cite{gpt4omini}.  The other three {\bf GPT-4}s are for ablation study in Sec~\ref{exp:ablation}.
  \item {\bf Llama-3}: Meta's {\bf Llama-3-8B-Instruct}~\cite{llama3} and {\bf Llama-3.1-8B-Instruct}~\cite{llama3dot1-paper}. The two larger {\bf Llama-3-70B-Instruct}~\cite{llama3-70b} and {\bf Llama-3.1-70B-Instruct} are for ablation.
  \item {\bf FLAN-T5}: Google's {\bf flan-t5-xl}~\cite{chung2024scaling}.
  \item {\bf Falcon}: TIIUAE's {\bf Falcon-7B-Instruct}~\cite{almazrouei2023falcon}.
  \item {\bf Vicuna-1.5}: LMSYS's {\bf vicuna-7b-v1.5}~\cite{chiang2023vicuna}.
  \item {\bf Mixtral}: Mistral AI's {\bf Mistral-7B-Instruct}~\cite{jiang2023mistral}.
  \item {\bf Phi-3}: Microsoft's {\bf Phi-3-mini-4k-instruct}~\cite{phi3}.
  \item {\bf Qwen2}: Alibaba's {\bf Qwen2-7B-Instruct}~\cite{yang2024qwen2}. The larger {\bf Qwen2-72B-Instruct}~\cite{qwen2-72b} is for ablation.
  \item {\bf LLaVA-1.6}: It is a large multimodal model (LMM). We evaluate on {\bf llava-v1.6-mistral-7b-hf}~\cite{liu2024visual,liu2024llavanext}.
  \item {\bf LLaRA}: It is domain-specific fine-tuned~\cite{liao2024llara} for ablation.
\end{itemize}

Note, LLaRA is the fine-tuned LlaMA-2-7B for recommender system domain.

\begin{figure}[bt!]
\centering
\begin{subfigure}{.24\textwidth}
  \centering
  \includegraphics[width=0.95\linewidth]{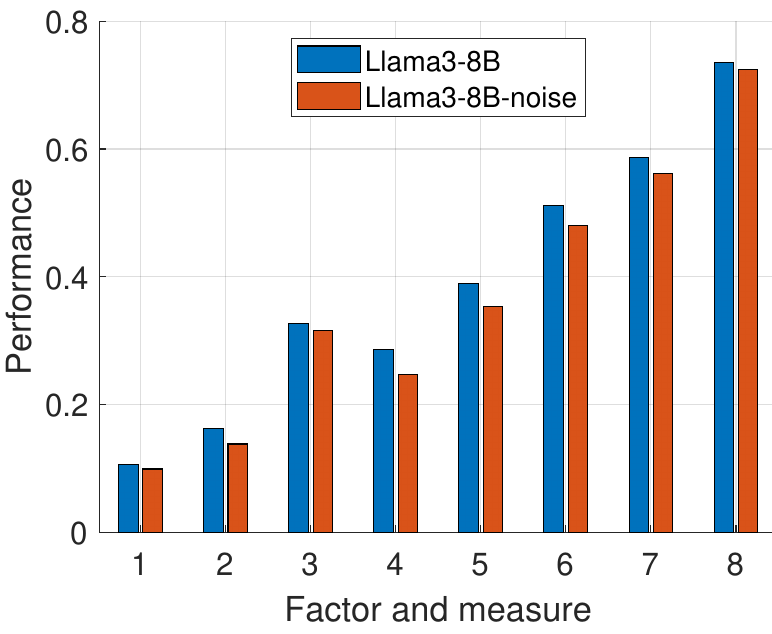} 
\end{subfigure}%
\begin{subfigure}{.24\textwidth}
  \centering
  \includegraphics[width=0.95\linewidth]{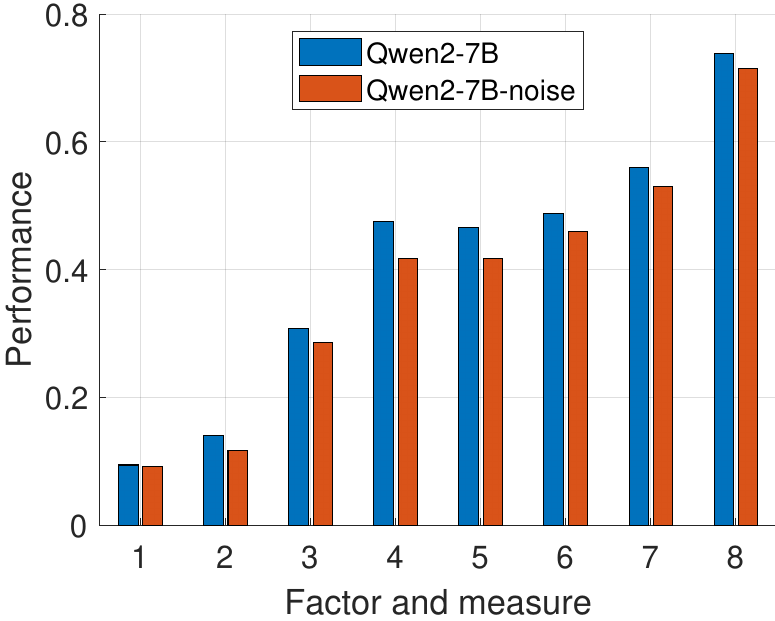}
\end{subfigure}%

\begin{subfigure}{.25\textwidth}
  \centering
  \includegraphics[width=0.95\linewidth]{materials/noisy_profiles_Llama3_70B.pdf}
\end{subfigure}
\caption{Robustness to Noisy and Fake Profiles.  (see Table~\ref{tab:noisy-profiles} for the index of x-axis and detailed values). (go back to Figure~\textcolor{blue}{\ref{fig:noisy-profiles}}) }
\label{fig:noisy-profiles-more}
\end{figure}

\subsection{Main Results on LRWorld (Netflix)}

The result on LRWorld (Netflix) is shown in Table~\ref{tab:Netflix}.

\subsection{More Results on Noisy Profiles to LLMs in RecSys}

See Figure~\ref{fig:noisy-profiles-more} for more results of LLMs' robustness to noisy and fake user profiles.

\subsection{Evaluation Prompts of the mini LRWorld (LastFM)}

The evaluation prompts are shown in Table~\ref{tab:prompts-lrworld-lastfm}.

\begin{table*}[tb!]
\caption{Evaluation Prompts of the Two Factors on the mini LRWorld (LastFM) dataset. (go back to Section~\textcolor{blue}{\ref{exp:lrworld-lastfm}}) }
\label{tab:prompts-lrworld-lastfm}
\resizebox{.95\textwidth}{!}{
\begin{tabular}{l|l}
\toprule
Factor  &  Prompt    \\    
\midrule  
<history, next-artist>   &   \begin{tabular}[c]{@{}l@{}}This user has listened to [HistoryHere] in the previous. \\ Please predict the next music artist this user will listen to. The artist name candidates are [CansHere]. \\ Choose only one artist from the candidates. The answer is  \end{tabular} \\ \midrule 
<artist, genre> & \begin{tabular}[c]{@{}l@{}}This user has listened to [HistoryHere] in the previous. \\ Please predict the genre of the next music artist this user will next listen to. The genres candidates are [CansHere]. \\ Choose only one genre from the candidates. The answer is   \end{tabular}   \\ 
\bottomrule                                                                                                                                                     
\end{tabular}
}
\end{table*}

\subsection{Inference Time of LLMs}

The inference time of LLMs on the largest LRWorld (Amazon) dataset is shown in Table~\ref{tab:Amazon-time}. The time format is HH:MM:SS (hour-minute-second). The implementation environment is as follows. IDE: PyCharm Community 2023, OS: Windows 10, Language: Python 3.10.13, CPUs: Intel Xeon E5-2673, Memory: 64GB, GPUs: Nvidia 3090. The time is calculated by the Python package tqdm. The open-source LLMs are the checkpoints from official versions on Hugging Face. The GPTs are called via OpenAI's APIs. Note, to avoid request rejection by the OpenAI's APIs, we execute time.sleep(seconds) per request in the program accordingly to monitor the progress. The inference time is only for reference since the Windows OS system is running other applications.

\subsection{Related Works on Evaluation-oriented LLMs in RecSys}\label{appendix:related-work}

There are lots of study and surveys on LLMs for RecSys~\cite{wu2023survey,wang2023generative,li2024survey}. We briefly review related works from the perspective of evaluation-oriented focus on LLMs in RecSys, as shown in Table~\ref{tab:related-work-list}. We categorize them by the evaluation factors and fields' richness of datasets. Their LLMs versions or backbones are also given. Note, the Fields column may not be complete  since their original papers might not clearly show what fields are used in the evaluated datasets.

\begin{table*}[tb!]
\caption{Effect of Sizes of LLMs on LRWorld (Amazon). The factor index is corresponding to the x-axis in Figure~\ref{fig:largerModels-Fewshot}. See Table~\ref{tab:scales-factors-description} for the detailed description of each factor. See Table~\ref{tab:model-version} for the detailed versions of the LLMs. Improvement 1: the relative improvement of the larger Llama3-70B model over its smaller Llama3-8B counterpart.  Improvement 2: the relative improvement of the larger Llama3.1-70B model over its smaller Llama3.1-8B  counterpart. Improvement 3: the relative improvement of the larger Qwen2-72B model over its smaller Qwen2-7B counterpart. (go back to Figure~\textcolor{blue}{\ref{fig:largerModels-Fewshot}})}
\label{tab:larger-model-improvement}
\resizebox{.94\textwidth}{!}{
\begin{tabular}{l|c|c|c|c|c|c|c|c|c|c|c}
\toprule
Factor Index & 1    & 2    & 3    & 4    & 5   & 6    & 7   & 8   & 9   & 10    & 11    \\
\midrule
Factor       & \multicolumn{1}{c|}{\begin{tabular}[c]{@{}c@{}}Neu emb\\ user2items\end{tabular}} & \multicolumn{1}{c|}{\begin{tabular}[c]{@{}c@{}}Neu emb\\ item2users\end{tabular}} & \multicolumn{1}{c|}{\begin{tabular}[c]{@{}c@{}}Mem \\ user-user\end{tabular}} & \multicolumn{1}{c|}{\begin{tabular}[c]{@{}c@{}}Taxo\\ \textless{}prod, root\textgreater{}\end{tabular}} & \multicolumn{1}{c|}{\begin{tabular}[c]{@{}c@{}}Multimodal\\ Knowledge\end{tabular}} & \multicolumn{1}{c|}{\begin{tabular}[c]{@{}c@{}}Mem\\ item-item\end{tabular}} & \multicolumn{1}{c|}{\begin{tabular}[c]{@{}c@{}}KG\\ \textless{}prod, color\textgreater{}\end{tabular}} & \multicolumn{1}{c|}{\begin{tabular}[c]{@{}c@{}}Taxo\\ \textless{}prod, leaf\textgreater{}\end{tabular}} & \multicolumn{1}{c|}{\begin{tabular}[c]{@{}c@{}}Taxo\\  \textless{}prod, interm\textgreater{}\end{tabular}} & \multicolumn{1}{c|}{\begin{tabular}[c]{@{}c@{}}Assoc\\ rule\end{tabular}} & \multicolumn{1}{c}{\begin{tabular}[c]{@{}c@{}}KG\\ \textless{}prod, brand\textgreater{}\end{tabular}} \\  
\midrule \midrule
Llama3-8B & 0.106 & 0.095 & 0.1618 & 0.3266 & 0.2861 & 0.554 & 0.3894 & 0.5122 & 0.5866 & 0.2372 & 0.7348 \\ \hline
Llama3.1-8B & 0.109 & 0.095 & 0.1508 & 0.3865 & 0.2897 & 0.5442 & 0.3948 & 0.5099 & 0.5966 & 0.5254 & 0.7506 \\ \hline
Qwen2-7B & 0.094 & 0.105 & 0.1404 & 0.3082 & 0.475 & 0.5114 & 0.4654 & 0.4877 & 0.5605 & 0.4067 & 0.7383 \\ \hline
Llama3-70B & 0.104 & 0.096 & 0.1785 & 0.3842 & 0.545 & 0.5639 & 0.4839 & 0.5145 & 0.6242 & 0.6694 & 0.7637 \\ \hline
Llama3.1-70B & 0.106 & 0.096 & 0.1646 & 0.3872 & 0.6031 & 0.5803 & 0.5299 & 0.5467 & 0.6334 & 0.661 & 0.7618 \\ \hline
Qwen2-72B & 0.096 & 0.095 & 0.1731 & 0.4831 & 0.5189 & 0.5311 & 0.5553 & 0.5582 & 0.6157 & 0.6525 & 0.7846 \\ 
\midrule \midrule
Improvement 1&	-1.89\% & 	1.05\% & 	10.32\% & 	17.64\% & 	90.49\% & 	1.79\% & 	24.27\% & 	0.45\% & 	6.41\% & 	182.21\% & 	3.93\% \\ \hline
Improvement 2& 	-2.75\% & 	1.05\% & 	9.15\% & 	0.18\% & 	108.18\% & 	6.63\% & 	34.22\% & 	7.22\% & 	6.17\% & 	25.81\% & 	1.49\% \\ \hline
Improvement 3&	2.13\% & 	-9.52\% & 	23.29\% & 	56.75\% & 	9.24\% & 	3.85\% & 	19.32\% & 	14.46\% & 	9.85\% & 	60.44\% & 	1.49\% \\ 
\bottomrule                                                                            
\end{tabular}
}
\end{table*}

\begin{table*}[tb!]
\caption{A Case Study on OpenAI's GPTs on the LRWorld (Netflix) Dataset: The Predictions. The similarity is scaled by x100 to see the score more clearly. See Figure~\ref{fig:error-gpts-explanation} for explanations. (go back to Table~\textcolor{blue}{\ref{tab:gpts}}) }
\label{tab:gpts-errors}
\resizebox{.8\textwidth}{!}{
\begin{tabular}{l|l|l}
\toprule
\multicolumn{2}{l}{\textbf{The target movie}:}                                                                                                                       \\ \midrule 
-  & \multicolumn{2}{l}{Training Day (2001),Dramas|Thrillers}       \\ 
\midrule  
\multicolumn{2}{l|}{\textbf{Which movie is the most similar to the target movie?}}          & \textbf{Item-Item Similarity to the Target Movie}                                                                                                          \\ \midrule 
1 & \begin{tabular}[c]{@{}l@{}}Patriot Games (1992),Action \& Adventure.\end{tabular}     & 25.15 \\ \hline
2 & \begin{tabular}[c]{@{}l@{}}Hum Aapke Hain Koun (1994),Classic Movies|Dramas|International Movies.\end{tabular}                &  0.19      \\ \hline
3 & \begin{tabular}[c]{@{}l@{}}The Time Machine (2002),Action \& Adventure|Sci-Fi \& Fantasy.\end{tabular}      &  19.94 \\ \hline
4 & \begin{tabular}[c]{@{}l@{}}Miracle (2004),Children \& Family Movies|Dramas|Sports Movies.\end{tabular}    &   21.41 \\ \hline
5 & \begin{tabular}[c]{@{}l@{}}Dil Hai Tumhaara (2002),Dramas|International Movies|Music \& Musicals.\end{tabular}    & 0.08 \\ \hline
6 & \begin{tabular}[c]{@{}l@{}}Butterfield 8 (1960),Classic Movies|Dramas|Romantic Movies.\end{tabular}             &      1.76     \\ \hline
7 & \begin{tabular}[c]{@{}l@{}}The Outlaw Josey Wales (1976),Action \& Adventure|Classic Movies.\end{tabular}    &  15.69  \\ \hline
8 & \begin{tabular}[c]{@{}l@{}}Red Dawn (1984),Action \& Adventure|Cult Movies.\end{tabular}    &  17.06 \\ \hline
9 & \begin{tabular}[c]{@{}l@{}}Shabd (2005),Dramas|International Movies|Romantic Movies.\end{tabular}     & 0.07 \\ \hline
10 (\textbf{Ground Truth}) & \begin{tabular}[c]{@{}l@{}}S.W.A.T. (2003),Action \& Adventure.\end{tabular}                    &  26.12   \\ \midrule  
\textbf{GPT-3.5-Turbo}      & \multicolumn{2}{l}{10 }            \\ \hline
\textbf{GPT-4o-mini}     & \multicolumn{2}{l}{9 $\quad$ (See Figure~\ref{fig:error-gpts-explanation} for explanations) }               \\ \hline
\textbf{GPT-4o}     & \multicolumn{2}{l}{9  }         \\ \hline
\textbf{GPT-4-Turbo}      & \multicolumn{2}{l}{10 }        \\ \hline
\textbf{GPT-4}     & \multicolumn{2}{l}{10 }     \\ 
\bottomrule                                                                                                                                                     
\end{tabular}
}
\end{table*}

\begin{figure*}[tb!]
    \centering
    \begin{subfigure}[t]{0.49\textwidth}
        \centering
        \includegraphics[width=0.96\linewidth]{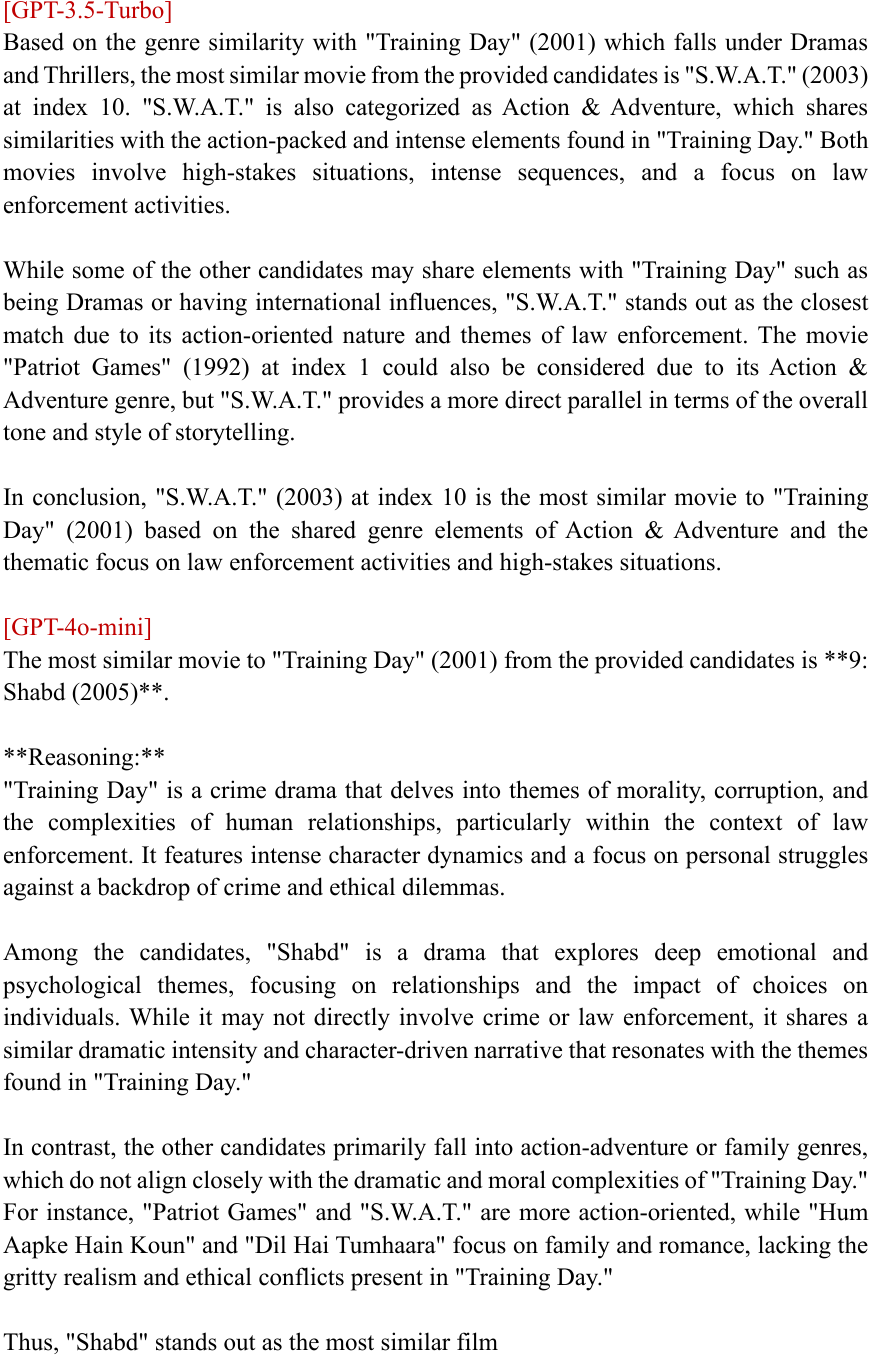}
    \end{subfigure}%
    \begin{subfigure}[t]{0.49\textwidth}
        \centering
        \includegraphics[width=0.925\linewidth]{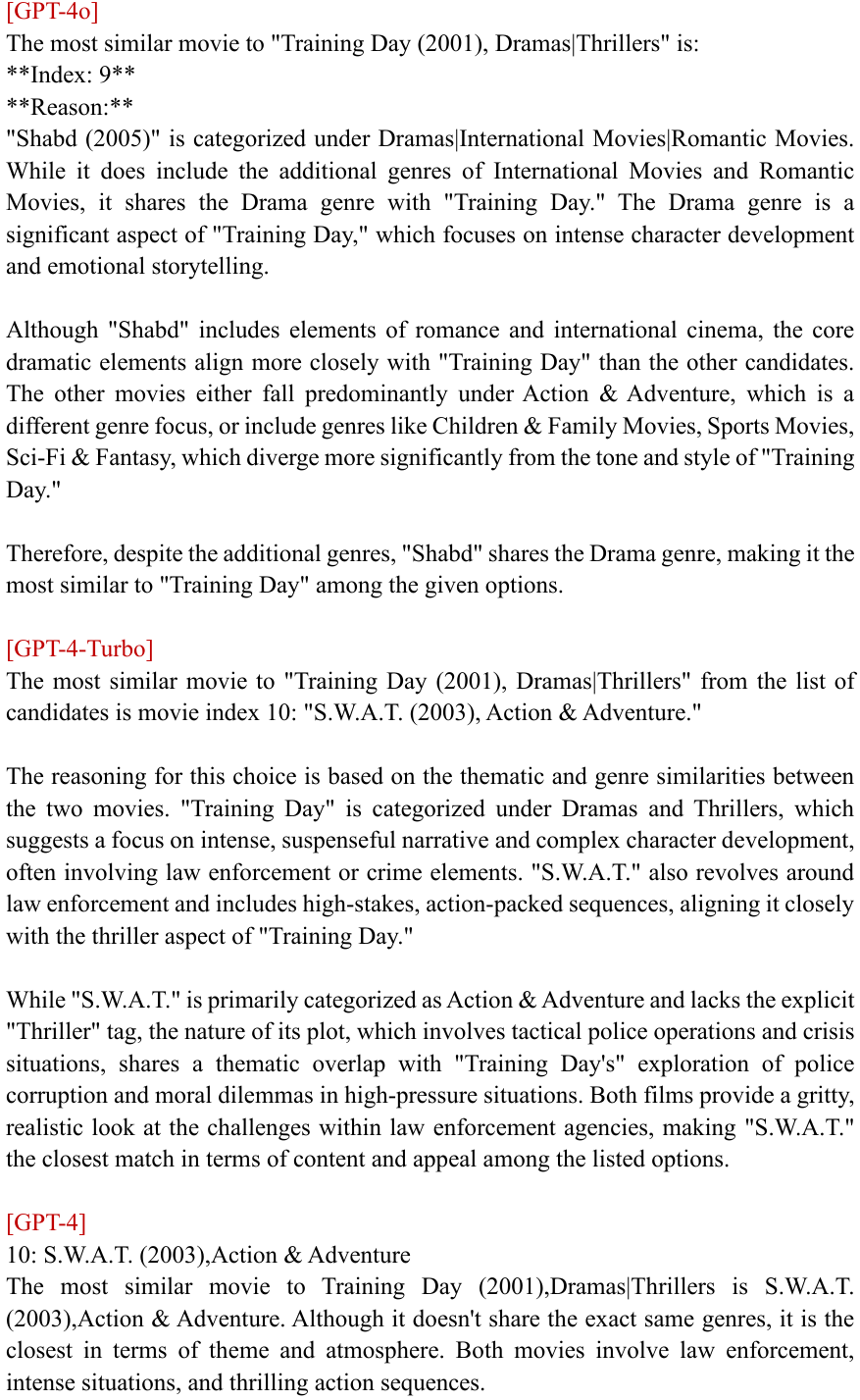}
    \end{subfigure}%
    \caption{A Case Study on OpenAI's GPTs on the LRWorld (Netflix) Dataset: The Reasons. Note, we cut the max new tokens returned by OpenAI's response at 256. See Table~\ref{tab:gpts-errors} for the context. We can see that, the behaviors of GPT-4o-mini and GPT-4o seem to be highly consistent.} 
\label{fig:error-gpts-explanation}
\end{figure*}

\begin{table*}[tb!]
\caption{A Brief Summary of Related Works from the Perspective of Evaluation on LLMs in RecSys. (go back to Section~\textcolor{blue}{\ref{paper:related-work}})}
\label{tab:related-work-list}
\resizebox{.99\textwidth}{!}{
\begin{tabular}{l|l|l|l|l|l}
\toprule
Work                 & Factors    & Fields      & Datasets   & LLMs     & Ref \\ 
\midrule
Zhang et   al, 2021  & \textless{}user-history, next-item\textgreater{}                                                                                                                                                         & user, item, title                                                                              & MovieLens-1M                                                                                                                     & GPT-2      &      \cite{zhang2021language}	\\ \hline
M6                   & \textless{}user-history,next-item\textgreater{}                                                                                                                                                          & \begin{tabular}[c]{@{}l@{}}user,item,review,title,\\category,attribute,behavior \end{tabular}                                              & \begin{tabular}[c]{@{}l@{}}AlipayQuery, \\ Amazon-Movie \& Cloth\end{tabular}                                                    & M6                                                                                                                            &      \cite{cui2022m6}	\\ \hline
P5                   & \begin{tabular}[c]{@{}l@{}}\textless{}user,item,rating\textgreater{},   \\ \textless{}user-history,next-item\textgreater{}\end{tabular}                                                                  & \begin{tabular}[c]{@{}l@{}}user,item,name,\\ title,review\end{tabular}                                                                         & \begin{tabular}[c]{@{}l@{}}Amazon Sports \& \\ Beauty \& Toys\end{tabular}                                                       & \begin{tabular}[c]{@{}l@{}}GPT-2, \\ T5\end{tabular}                                                                          &      \cite{geng2022recommendation}	\\ \hline
BIGRec               & \textless{}user-history,next-item\textgreater{}                                                                                                                                                          & user,item,title                                                                                                                              & \begin{tabular}[c]{@{}l@{}}MovieLens10M, \\ Amazon Game\end{tabular}                                                             & LLaMA-7B                                                                                                                      &      \cite{bao2023bi}	\\ \hline
TransRec    & \textless{}user-history,next-item\textgreater{}                                                                                                                                                          & user,item,title,attribute                                                                                                                    & \begin{tabular}[c]{@{}l@{}}Amazon Beauty \& Toys, \\ Yelp\end{tabular}                                                           & \begin{tabular}[c]{@{}l@{}}P5, BART-large, \\ LLaMA-7B\end{tabular}                                                        &      \cite{lin2024bridge}	\\ \hline
Chat-REC             & \begin{tabular}[c]{@{}l@{}}\textless{}user,item,rating\textgreater{},   \\ \textless{}user-history,next-item\textgreater{}\end{tabular}                                                                  & user,item,title,demographics                                                                                                                   & MovieLens 100K                                                                                                                   & gpt-3.5-turbo                                                                                                                 &      \cite{gao2023chat}	\\ \hline
CoLLM                & \textless{}user-history,next-item\textgreater{}                                                                                                                                                          & user,item,title                                                                                                                                & \begin{tabular}[c]{@{}l@{}}MovieLens-1M, \\ Amazon Book\end{tabular}                                                             & Vicuna-7B                                                                                                                     &      \cite{zhang2023collm}	\\ \hline
Kang et   al, 2023   & \textless{}user, item, rating\textgreater{}                                                                                                                                                              & user, item, genre, brand, title                                                                                                                & \begin{tabular}[c]{@{}l@{}}MovieLens-1M, \\ Amazon-Book\end{tabular}                                                             & \begin{tabular}[c]{@{}l@{}}gpt-3.5-turbo, \\ Flan-U-PaLM/-T5\end{tabular} &      \cite{kang2023llms}	\\ \hline
E4SRec               & \textless{}user-history,next-item\textgreater{}                                                                                                                                                          & user,item                                                                                                                                      & \begin{tabular}[c]{@{}l@{}}Amazon Beauty, Sports \& Toy,   \\ Yelp\end{tabular}                                                  & LLaMA2-13B                                                                                                                    &      \cite{li2023e4srec}	\\ \hline
GPTRec & \textless{}user-history, next-K-item\textgreater{}                                                                                                                                                       & user,item                                                                                                                                      & MovieLens-1M                                                                                                                     & GPT-2                                                                                                                         &      \cite{petrov2023generative}	\\ \hline
Liu et   al, 2023    & \begin{tabular}[c]{@{}l@{}}\textless{}user,item,rat\textgreater{},   \\ \textless{}user-history,next-item\textgreater{}, \\ \textless{}expl\textgreater{}, \textless{}summary\textgreater{}\end{tabular} & user, item, title,  category                                                                                                                 & Amazon Beauty                                                                                                                    & \begin{tabular}[c]{@{}l@{}}gpt-3.5-turbo, \\ T0, P5, GPT-2\end{tabular}                                                 &      \cite{liu2023chatgpt}	\\ \hline
FaiRLLM  & fairness                                                                                                                                                                                  & \begin{tabular}[c]{@{}l@{}}singer,item,director,demographics \end{tabular} & Music, Movie                                                                                                                     & gpt-3.5-turbo                                                                                                                 &      \cite{zhang2023chatgpt}	\\ \hline
Sanner   et al, 2023 & \textless{}user-history, next-item\textgreater{}                                                                                                                                                         & \begin{tabular}[c]{@{}l@{}}user, item, \\ desc\end{tabular}                                                                                    & \begin{tabular}[c]{@{}l@{}}Amazon Applianc \& Gift Cards \\  \& Prime Pantry \& Software\end{tabular}                         & PaLM                                                                                                                          &      \cite{sanner2023large}	\\ \hline
He et al,   2023     & \textless{}user-history,next-item\textgreater{}                                                                                                                                                          & \begin{tabular}[c]{@{}l@{}}user,item,title, cast, \\ genre, utterance\end{tabular}                                                          & \begin{tabular}[c]{@{}l@{}}INSPIRED, ReDIAL, \\ Reddit-Movie\end{tabular}                                                     & \begin{tabular}[c]{@{}l@{}}GPT-3.5-turbo, GPT-4, \\ Vicuna, baize-v2-13b\end{tabular}                                &      \cite{he2023large}	\\ \hline
Harte et   al, 2023  & \textless{}user-history,next-item\textgreater{}                                                                                                                                                          & \begin{tabular}[c]{@{}l@{}}user,item,name, session\end{tabular}                                                                              & \begin{tabular}[c]{@{}l@{}}Amazon Beauty, Delivery Hero\end{tabular}                                                          & OpenAI ada                                              &      \cite{harte2023leveraging}	\\ \hline
LlamaRec             & \textless{}user-history,next-item\textgreater{}                                                                                                                                                          & user,item,title,                                                                                                                               & \begin{tabular}[c]{@{}l@{}}ML-100k, Amazon Beauty \&  Games\end{tabular}                                                      & Llama 2-7B                                                                                                                    &      \cite{yue2023llamarec}	\\ \hline
POD                  & \textless{}user-history,next-item\textgreater{}                                                                                                                                                          & user,item,review                                                                                                                               & Amazon Sports \& Beauty \& Toys                                                                                                  & T5-small                                                                                                                      &      \cite{li2023prompt}	\\ \hline
InstructRec          & \textless{}user-history,next-item\textgreater{}                                                                                                                                                          & \begin{tabular}[c]{@{}l@{}}user, item, title,  category\end{tabular}                                                                       & Amazon Games \& CDs                                                                                                              & \begin{tabular}[c]{@{}l@{}}text-davinci-003,   \\ Flan-T5-XL\end{tabular}                                           &      \cite{zhang2023recommendation}	\\ \hline
iEvaLM   & conversations                                                                                                                                                                                            & \begin{tabular}[c]{@{}l@{}}user, item,  utterance\end{tabular}                                                                               & \begin{tabular}[c]{@{}l@{}}ReDial, OpenDialKG\end{tabular}                                                                    & gpt-3.5-turbo                                                                                                                 &      \cite{wang2023rethinking}	\\ \hline
TALLRec              & \textless{}user-history, next-item\textgreater{}                                                                                                                                                         & \begin{tabular}[c]{@{}l@{}}user,item,title,director,desc\end{tabular}                                                                       & \begin{tabular}[c]{@{}l@{}}MovieLens100K, BookCrossing\end{tabular}                                                           & LLaMA-7B                                                                                                                      &      \cite{bao2023tallrec}	\\ \hline
Dai et   al, 2023    & \begin{tabular}[c]{@{}l@{}}\textless{}user,item,rating\textgreater{},   \\ \textless{}user-history,next-item\textgreater{}\end{tabular}                                                                  & user,item,title                                                                                                                                & \begin{tabular}[c]{@{}l@{}}MovieLens-1M, MIND-small \\ Amazon Books \&  Music \\ \end{tabular}                                  & \begin{tabular}[c]{@{}l@{}}gpt-3.5-turbo, \\ text-davinci-003 \end{tabular}                             &      \cite{dai2023uncovering}	\\ \hline
Visual P5            & \begin{tabular}[c]{@{}l@{}}\textless{}user-history,next-item\textgreater{},   \\ \textless{}user, matched-item\textgreater{}\end{tabular}                                                                & \begin{tabular}[c]{@{}l@{}}user,item,review,desc,img\end{tabular}                                                                           & \begin{tabular}[c]{@{}l@{}}Amazon Clothing, Shoes \&  Jewelry,  Beauty,  \\ Sports \& Outdoors, Toys, \& Games\end{tabular} & P5-small                                                                                                                      &      \cite{geng2023vip5}	\\ \hline
Cao et   al, 2024    & \begin{tabular}[c]{@{}l@{}}\textless{}user,item,rating\textgreater{},   \\ \textless{}user-history,next-item\textgreater{}\end{tabular}                                                                  & user,item,title                                                                                                                                & \begin{tabular}[c]{@{}l@{}}Amazon Toys \& Games, Beauty,  \\ and Sports \& Outdoors\end{tabular}                                 & \begin{tabular}[c]{@{}l@{}}FLANT5-Base, \\ FLAN-T5-XL\end{tabular}                                                            &      \cite{cao2024aligning}	\\ \hline
CoRAL                & \textless{}user-history,next-item\textgreater{}                                                                                                                                                          & user,item,title                                                                                                                                & \begin{tabular}[c]{@{}l@{}}Amazon Applianc \& Gift Cards  \\ \& Prime Pantry \& Software\end{tabular}                            & \begin{tabular}[c]{@{}l@{}}PaLM, \\ GPT-4\end{tabular}                                                                        &      \cite{wu2024coral}	\\ \hline
DEALRec              & \textless{}user-history,next-item\textgreater{}                                                                                                                                                          & user,item,title,                                                                                                                               & \begin{tabular}[c]{@{}l@{}}Amazon Games \& Book,\\ MicroLens-50K\end{tabular}                                                    & LLaMA-7B                                                                                                                      &      \cite{lin2024data}	\\ \hline
Palma et   al, 2024  & \textless{}user-history, next-item\textgreater{}                                                                                                                                                         & user, item, genre, author                                                                                                                      & \begin{tabular}[c]{@{}l@{}}MovieLens, LastFM, FB  Books\end{tabular}                                                       & \begin{tabular}[c]{@{}l@{}}ChatGPT-3.5, \\ ChatGPT-4\end{tabular}                                                             &      \cite{di2023evaluating}	\\ \hline
GenRec               & \textless{}user-history,  next-K-item\textgreater{}                                                                                                                                                      & user,item,title                                                                                                                                & \begin{tabular}[c]{@{}l@{}}MovieLens-25M, Amazon Toys\end{tabular}                                                            & Llama-7b                                                                                                                      &      \cite{ji2024genrec}	\\ \hline
STELLA               & \textless{}user-history,next-item\textgreater{}                                                                                                                                                          & user,item,title                                                                                                                               & \begin{tabular}[c]{@{}l@{}}MovieLens-1M, MIND-small, \\ Amazon Books \& Music \end{tabular}                                   & GPT-3.5-turbo                                                                                                                 &      \cite{ma2023large}	\\ \hline
Hou et al, 2024      & seq rec, retrieval, ranking                                                                                                                                                                              & user, item, title                                                                                                                             & \begin{tabular}[c]{@{}l@{}}ML-1M,   Amazon Games\end{tabular}                                                                   & gpt-3.5-turbo                                                                                                                 &      \cite{hou2024large}	\\ \hline
PeaPOD               & \begin{tabular}[c]{@{}l@{}}\textless{}user-history,next-item\textgreater{},   \\ \textless{}expl\textgreater{}\end{tabular}                                                                              & user,item                                                                                                                                      & Amazon Sports \& Beauty \& Toys                                                                                                  & \begin{tabular}[c]{@{}l@{}}T5-small, P5, Visual P5\end{tabular}                                                         &      \cite{ramos2024preference}	\\ \hline
LLMRec               & \begin{tabular}[c]{@{}l@{}}\textless{}user-history,next-item\textgreater{}, <expl>  \\  <user,item,rat>,  <summary> \end{tabular}                                                                              & user,item,review,attribute      & Amazon Beauty                                                                                & \begin{tabular}[c]{@{}l@{}}gpt-3.5-turbo, LLaMa-7B, \\ ChatGLM-6B, Alpaca\end{tabular}                                                         &      \cite{liu2023llmrec}	\\ \hline
ReLLa                & \textless{}user-history,next-item\textgreater{}                                                                                                                                                          & \begin{tabular}[c]{@{}l@{}}user,item,title,genre,demographics\end{tabular}                                                            & \begin{tabular}[c]{@{}l@{}}MovieLens-1M/-25M,   BookCrossing\end{tabular}                                        & \begin{tabular}[c]{@{}l@{}}Vicuna-7B/-13B\end{tabular}                                                              &      \cite{lin2024rella}	\\ \hline
TokenRec             & \textless{}user-history, next-item\textgreater{}                                                                                                                                                         & user,item,title                                                                                                                                & \begin{tabular}[c]{@{}l@{}}MovieLens-1M, LastFM \\ Amazon Beauty \& Clothing \end{tabular}                                    & T5-small                                                                                                                      &      \cite{qu2024tokenrec}	\\ \hline
Xu et al,   2024     & \textless{}user-history,next-item\textgreater{}                                                                                                                                                          & user,item,title                                                                                                                                & \begin{tabular}[c]{@{}l@{}}MovieLens-1M, \\ Yelp,   Amazon  Beauty\end{tabular}                                                 & {\color[HTML]{1F2328} \begin{tabular}[c]{@{}l@{}}T5,   Llama2-7B\end{tabular}}                                             &      \cite{xu2024fairly}	\\ \hline
RDRec                & \textless{}user-history,next-item\textgreater{}                                                                                                                                                          & user,item,review,attribute                                                                                                                     & \begin{tabular}[c]{@{}l@{}}Amazon Sports \& Outdoors,   \\ Beauty,  Toys, \& Games\end{tabular}                                  & \begin{tabular}[c]{@{}l@{}}T5-small, \\ Llama-2-7b\end{tabular}                                                               &      \cite{wang2024rdrec}	\\ \hline
LLaRA                & \textless{}user-history,next-item\textgreater{}                                                                                                                                                          & user,item,title,artist                                                                                                                     & \begin{tabular}[c]{@{}l@{}}MovieLens100K, Steam, LastFM \end{tabular}   & \begin{tabular}[c]{@{}l@{}} Llama-2-7b\end{tabular}                                                               &      \cite{liao2024llara}	\\ \hline
ProLLM4Rec                & \textless{}user-history,next-item\textgreater{}                                                                                                                                                          & \begin{tabular}[c]{@{}l@{}}user,item,title,genre,\\brand,category,price,desc\end{tabular}                                                                                                                     & \begin{tabular}[c]{@{}l@{}}MovieLens-1M, Amazon-Books\end{tabular}   & \begin{tabular}[c]{@{}l@{}}ChatGLM/2/3,Flan-T5,LLaMA/2,\\ Vicuna,GPT-4,ChatGPT(2023)\end{tabular}     &      \cite{xu2024prompting}	\\ 
\bottomrule
\end{tabular}
}
\end{table*}

\end{document}